# Shape Ultrasound with Dynamic Microfluidic Lenses

*Dajun Zhang[1], Hyunji Shim[1], Michael Ferolito[1], Feng Ye[1], Hongrui Jiang[1], and Chu Ma[1]**

*[1]Department of Electrical and Computer Engineering, University of Wisconsin–Madison, Madison, WI 53706, USA*

**E-mail: chu.ma@wisc.edu*

## Abstract

Dynamic shaping of ultrasound into prescribed spatial patterns underlies a broad range of biomedical and engineering applications. However, existing modulation strategies face fundamental limitations: single element transducers paired with acoustic lenses lack reconfigurability, whereas phased arrays require large numbers of independently driven elements, leading to substantial hardware complexity, cost, and rigidity. Here we introduce a microfluidic ultrasound lens system that enables reconfigurable spatial modulation of ultrasonic fields using two orthogonal layers of soft microfluidic channels. Each channel is selectively filled with one of two liquids with distinct sound speeds, such as water and silicone oil, via an FPGA controlled array of micropumps, generating programmable binary phase patterns. Integrating a 20-row by 20-column microfluidic lens with a single element transducer, we demonstrate three-dimensional ultrasound focusing with approximately 1 s reconfiguration time and spatial resolution comparable to that of a 400-element transducer array. The system provides 400 addressable pixels through parallel control of 80 pumps, allowing hardware complexity to scale with the square root of the pixel count. Building on this platform, we demonstrate dynamic ultrasound heating with reconfigurable spatial heating patterns, as well as remote particle manipulation. By integrating the microfluidic lens with static holographic lenses, we achieve dynamically reconfigurable ultrasound fields with enhanced spatial control. Furthermore, we demonstrate the versatile design capabilities of the microfluidic lens platform through a cylindrical lens that manipulates ultrasound propagation in the azimuthal direction. Owing to its liquid based, soft architecture, the microfluidic lens offers design flexibility, scalable operation across ultrasound frequencies, low acoustic transmission loss, and stable performance under high acoustic power. Together, these results establish microfluidic phase modulation as a compact, scalable, and flexible approach for dynamic ultrasound field control.

Ultrasound can penetrate a wide range of materials and deliver acoustic energy in a non-invasive manner. The ability to precisely modulate and shape ultrasound into desired spatial patterns is becoming increasingly important, enabling advances in medical diagnostics, therapeutic interventions, and a growing number of non-medical applications[1–3]. Spatially modulated ultrasound enables thermal and mechanical interventions such as localized tumor ablation[4,5], non-invasive neuromodulation[6–8], cardiovascular therapies[9,10], heart pacemaking[11], targeted blood-brain-barrier opening[12–14], drug delivery[15–17], kidney-stone fragmentation[18], and cosmetic treatments[19]. Beyond biomedicine, spatially modulated ultrasound is being explored for object manipulation[20,21], acoustic volumetric printing[22,23], wireless power transfer[24], and holographic display[25]. Despite these broad opportunities, practical generation of adaptable, high-resolution ultrasound fields remains technically challenging.

Existing ultrasound modulation systems can be broadly classified into two categories: single-element transducer equipped with an additional lens, and multi-element transducer array. Lens-based systems shape wavefronts through tailored geometries or material properties, but most lenses are fixed after fabrication and offer little or no reconfigurability. Mechanical scanning can partially compensate for this inflexibility, yet inevitably leads to bulky platforms with slow refresh rates and limited accessibility in confined or wearable settings. By contrast, multi-element arrays enable dynamic modulation through independent control of element amplitudes and delays. However, ultrasound field synthesis typically requires hundreds to thousands of independently driven channels, causing rapid escalation in hardware complexity, cost and power consumption. Moreover, such systems are often bulky and mechanically rigid, making them difficult to conform to curved geometries, including most biological surfaces and many non-biological objects. Recent advances in flexible electronics have enabled soft, wearable ultrasound arrays[26,27] that mitigate some of these limitations by reducing bulkiness and improving conformability. However, current demonstrations are largely restricted to imaging, where acoustic power requirements are comparatively low. Delivering the substantially higher acoustic intensities needed for mechanical and thermal interventions, while maintaining high-density spatial control, remains a major unsolved challenge, particularly for soft and wearable platforms. Together, these limitations point to a critical need for dynamically reconfigurable ultrasound lenses that can provide array-like spatial control while avoiding the complexity, rigidity and power-handling constraints of existing systems.

In optics, dynamic spatial control is well established using spatial light modulators, which provide high pixel counts, rapid refresh rates and reconfigurability[28]. These devices have transformed optical wavefront engineering by enabling programmable focusing, beam shaping and holographic field generation. Achieving an acoustic analogue with comparable performance has proven more challenging because of the longer wavelengths and lack of effective ways to dynamically modulate acoustic properties of materials. Mechanical approaches to dynamic acoustic field control have been demonstrated[29–32], but are largely confined to low frequencies (kHz range) and typically support fewer than one hundred addressable pixels, because of the difficulty of scaling mechanical actuation to ultrasonic frequencies and larger pixel counts. Multiplexed holograms enable partial dynamic control by varying the incident wave frequency, wavefront, or surrounding medium, yet the accessible field patterns remain constrained by the underlying static[33,34] or quasi-static[35] hologram. Another approach employs a CMOS chip to generate on-demand microbubble patterns that form binary amplitude holograms[36]; however, serial pixel addressing and a mechanical wiping step limit the achievable refresh rate, and the strong ultrasound scattering by microbubbles and electronic components results in substantial acoustic energy loss. Notably, most existing dynamic ultrasound lenses are rigid and planar, restricting their applicability on curved, deformable or anatomically complex surfaces[29–36].

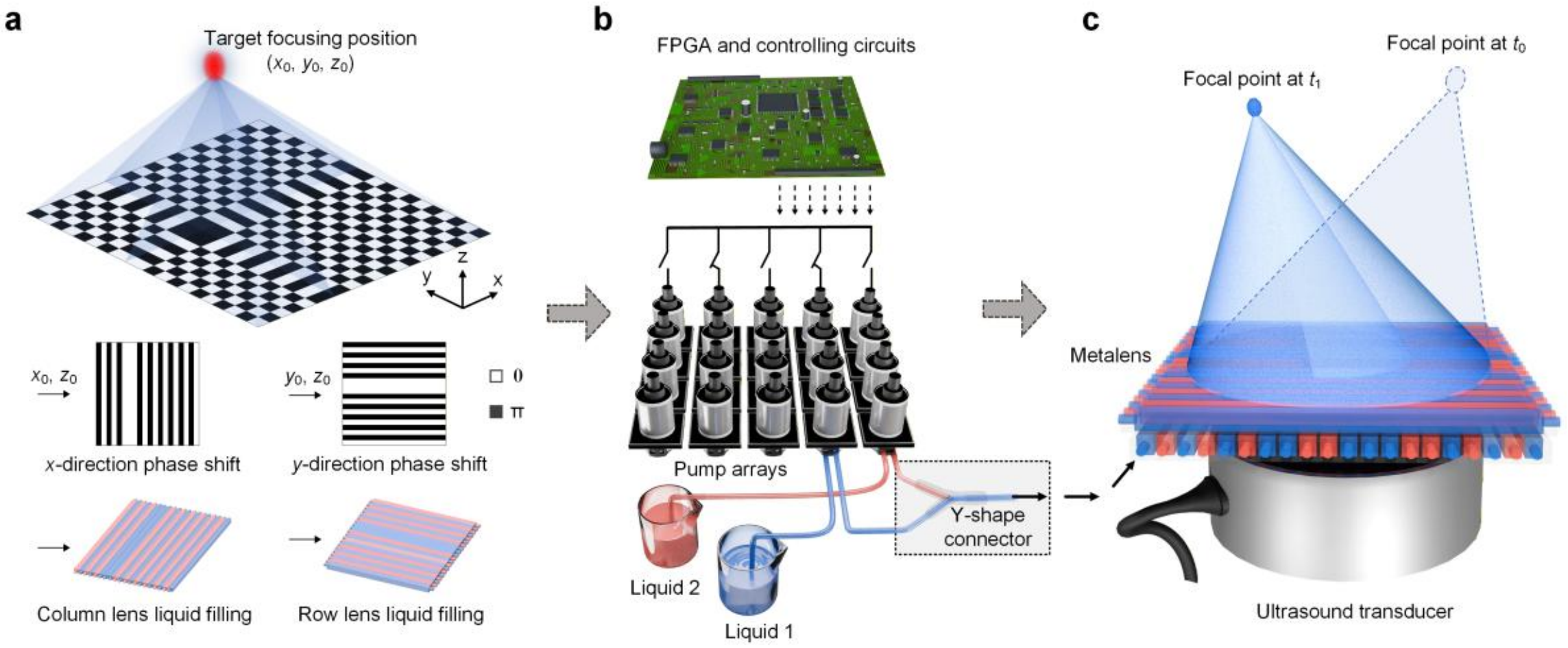

**Fig. 1 | Overview of the dynamic microfluidic ultrasound lens.** a. Schematic of the ultrasound shaping mechanism. Spatial focusing of ultrasound waves is achieved using two orthogonal microfluidic layers, a row layer and a column layer. Each layer consists of parallel 1D microfluidic channels selectively filled with one of two liquids, producing binary phase patterns corresponding to 0 and $\pi$ phase shifts. b. Schematic of the FPGA-controlled liquid-filling system. Each microfluidic channel is connected to two micropumps that deliver the two different liquids, and all pumps are switched in parallel through a multichannel FPGA controller. c. Integration of a single-element transducer with the microfluidic lens. A single-element transducer generates a uniform plane wave that enters the microfluidic lens. The transmitted ultrasound is then focused to user-defined spatial locations, reconfigurable through different liquid-filling patterns.

While acoustic waves have long been used within microfluidic systems, for example, in acoustic streaming and tweezing for particle manipulation and separation[37–39], the potential of microfluidics for acoustic wavefront engineering has yet to be explored. Here we report a soft, reconfigurable microfluidic ultrasound lens system that enables dynamic spatial modulation of ultrasonic fields without relying on multi-element transducer arrays. The system consists of orthogonally arranged row and column microfluidic lenses incorporating liquid channels (Fig. 1). By selectively filling microfluidic channels with two liquids of contrasting sound speeds, e.g., water and silicone oil in our demonstrations, using individually addressable, FPGA-controlled pumps, we generate programmable acoustic phase shifts across the aperture. Combining a single-element transducer and a 20-row × 20-column lens architecture, this approach enables three-dimensional ultrasound focusing with a reconfiguration time on the order of one second and a spatial resolution comparable to that of a 400-element transducer array. Importantly, 400 addressable acoustic pixels are realized through parallel control of only 80 pumps, such that system complexity and cost scale as $\sqrt{N}$ rather than $N$, in sharp contrast to conventional arrays requiring independent electronic control of every element. The liquid-based microfluidic architecture offers substantial design flexibility: channel dimensions can be readily scaled to operate across a wide range of ultrasound frequencies, and lens geometries can be fabricated in cylindrical, circular or arbitrary shapes owing to the soft and deformable nature of the materials. Because the pumps and control electronics are physically decoupled from the acoustic aperture, and thus removed from the acoustic path, wave modulation remains unaffected by the acoustic scattering from electronic components. Furthermore, the system exhibits strong thermal stability and robust operation under high ultrasonic power, making it well suited for applications requiring both high acoustic intensity and dynamic spatial control.

Phase modulation in the microfluidic lens is achieved by filling its fluidic channels with different liquids (e.g., water and silicone oil) with distinct acoustic velocities (see Supplementary Note 3). In the microfluidic lens, the row channels modulate the phase profile along the $x$-direction, while the column channels modulate it along the $y$-direction. Together, the two layers form a 2D matrix of binary phase pixels capable of generating a single focal spot, multiple foci, or more complex spatial acoustic patterns in the 3D region in front of the lens. For a target focal point $(x_0, y_0, z_0)$, the row layer produces a focal line aligned with $(y, z) = (y_0, z_0)$, and the column layer produces a focal line aligned with $(x, z) = (x_0, z_0)$. The binary phase patterns required for both layers, each corresponding to a specific liquid-filling configuration, are independently computed using an iterative optimization algorithm described in Supplementary Note 4.

Once the liquid pattern is determined, it is executed by an FPGA-controlled micropump array. Each microfluidic channel is connected to two micropumps through a Y-shaped junction, enabling selective filling with either water or silicone oil (Supplementary Fig. S4b and Supplementary Video 3). The superposition of the row-layer and column-layer phase profiles generate a 2D array of binary phase pixels that focuses ultrasound to a specified 3D target. To generate a different focusing pattern, the system activates the corresponding micropumps to reconfigure the liquid distribution within the channels. This enables rapid and repeatable reprogramming of the lens for dynamic ultrasound shaping.

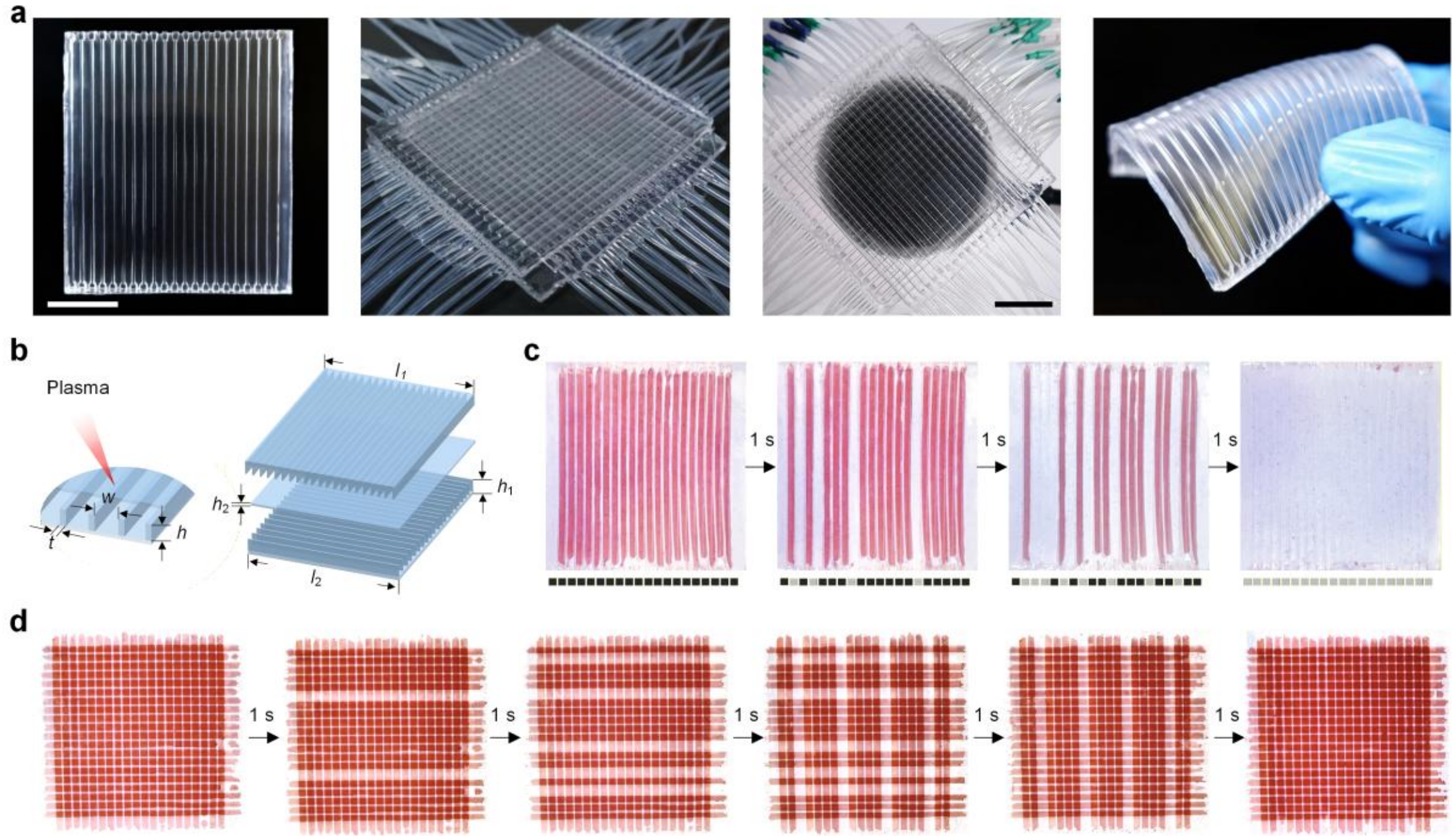

**Fig. 2 | Microfluidic lens design and liquid filling pattern reconfiguration. a.** Pictures of the PDMS-based microfluidic lens. Scale bars, 2mm. **b.** Details of the lens that works for 1 MHz ultrasound. $t$ =0.5 mm, $W$ = 2.5 mm, $h$ =1.5 mm, $h_2$ =0.5 mm, $l_1$ =60 mm, $l_2$= 70 mm. **c.** and **d.** Snapshots of liquid pattern reconfiguration process in a single-layer lens (c, Supplementary Video 1) and a double-layer row-column lens (d, Supplementary Video 2). The red liquid is water with red ink, and the transparent liquid is silicone oil.

Figure 2a shows the design and geometry of the microfluidic lens. The microfluidic lens is fabricated using PDMS. The incident ultrasound field is generated by a single-element transducer with a diameter of 6 cm and a center frequency of 1 MHz, corresponding to a wavelength of 1.5 mm in water. To generate a $\pi$ phase difference between the transmitted waves through water (sound speed $c_w$ = 1500 m/s) and silicone oil (sound speed $c_w$ = ~980 m/s), the thickness of the liquid channel is calculated as $h$ = 1.5 mm. The period of the channels is set as 3 mm, and the thickness of the PDMS walls is 0.5 mm. There are 20 row channels and 20 column channels, all having an effective length of 6 cm, giving a total effective lens size of 6 cm × 6 cm. The detailed fabrication procedure is described in the Supplementary Note 1.

## Ultrasound focusing pattern characterization

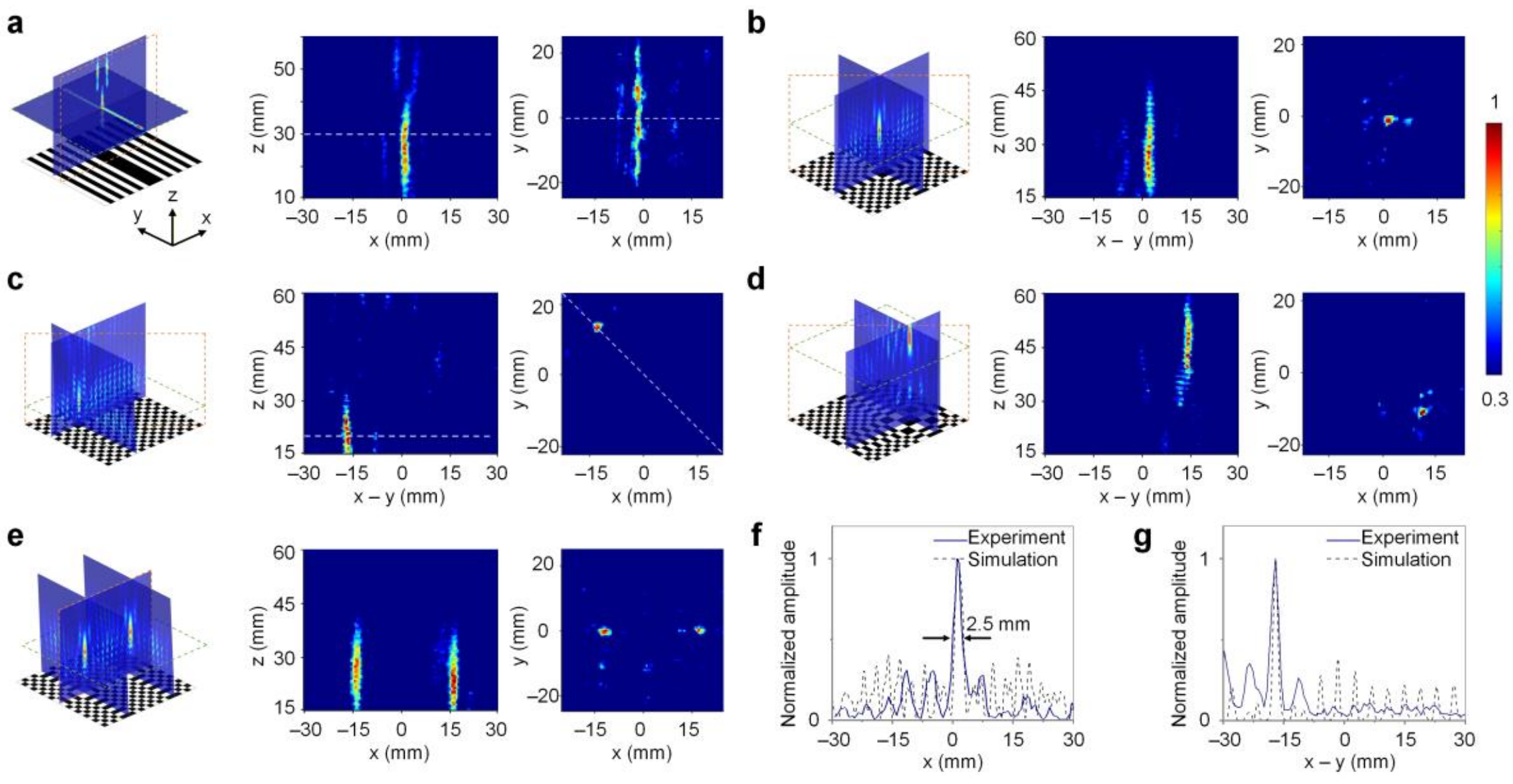

**Fig. 3 | Simulated and experimentally measured ultrasound focusing patterns by the microfluidic lens.** In a-e, the left plot shows simulation result, and the middle and right ones show experimentally measured field distributions. **a.** Simulated and experimentally measured focal line produced by the single layer lens using the liquid filling pattern "010101010111010101 01", where "0" denotes water and "1" denotes silicone oil. **b-e.** Simulated and experimentally measured focal patterns generated by the double-layer row-column lens: a single focal spot at ($x$, $y$, $z$) = (1.5 mm, 1.5 mm, 30 mm) (b), (–12 mm, –12 mm, 20 mm) (c), and (10.5 mm, 10.5 mm, 50 mm ) (d), and double focal spots at (–14.5 mm, –1.5 mm, 30 mm) and (16.5 mm, –1.5 mm, 30 mm) (e). **f.** Acoustic amplitude along the cut line at $z$ = 30 mm and $y$ =0 mm from the focusing pattern shown in (a). **g.** Acoustic amplitude along the cut line at $z$ = 20 mm and $x + y = 0$ from the focusing pattern shown in (c). The corresponding cut lines are indicated as dashed lines in (a) and (c).

We experimentally characterized the ultrasound field after the wave emitted by the single-element transducer is transmitted through the microfluidic lens, by scanning a needle transducer across planes perpendicular and parallel to the lens surface, passing through the expected focal regions. As shown in Fig. 3, a row or column layer generates well-defined line foci, and the row-column lens successfully focuses ultrasound to prescribed 3D spatial locations. The measured focal spot sizes and focal positions show good agreement with numerical simulations of the ultrasound field in MATLAB. Details of the numerical modeling and experimental measurement procedures are included in Supplementary Note 4 and 7. More examples of focal patterns can be found in Supplementary Note 4.3.

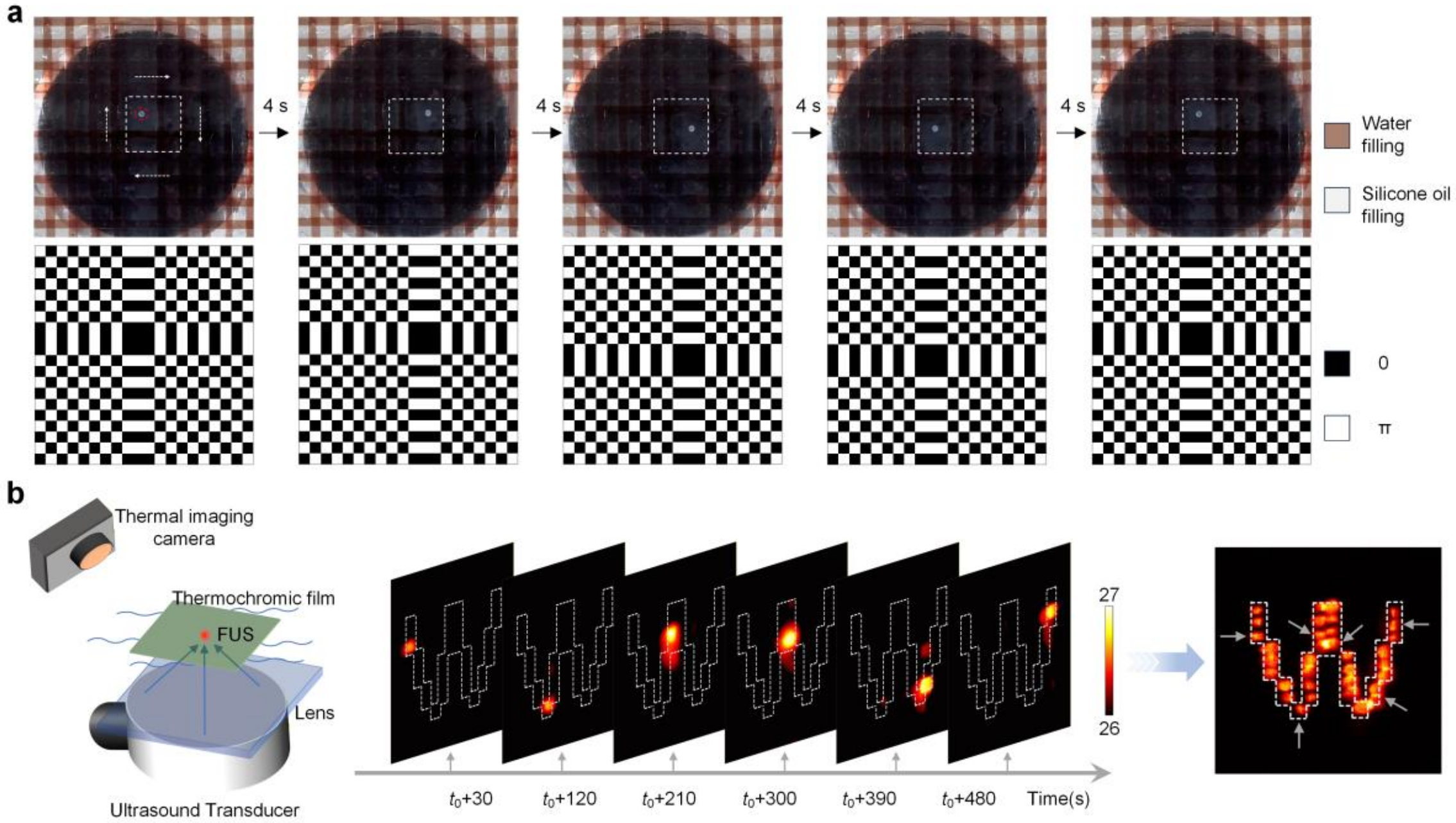

**Fig. 4 | Dynamic ultrasound shaping experiments. a.** Particle manipulation. Snapshots of Supplementary Video 4 showing four liquid filling patterns (bottom) designed to move a particle along a square trajectory (top). The particle is made of silicone rubber in a cylindrical shape and is floating on water surface. **b.** Spatially selective heating. Left: schematic of the experimental setup for spatially selective heating. A thermochromic film is placed 3 cm above the lens. Right: by reconfiguring the liquid filling patterns, all the locations along a letter "W" can be sequentially heated up with high spatial precision.

## Dynamic ultrasound shaping experiments

We conducted two representative demonstrations, particle manipulation and spatially selective heating, to illustrate the application potential enabled by the reconfigurable phase-modulation capability of the microfluidic system.

In the particle manipulation experiment (Fig. 4a), a cylindrical silicone-rubber particle (1.5 mm diameter, 1 mm height) was floated on the water surface, with the lens positioned 3 cm below. Four liquid filling configurations were designed for the microfluidic channels, each producing a focal spot at one corner of a square region on the water surface. As the filling patterns switched sequentially, the acoustic focus moved accordingly, driving the particle to levitate successively at the four corners of the square.

In the spatially selective heating experiment (Fig. 4b), a thermochromic film was positioned 3 cm from the lens, with the film, lens, and transducer all immersed in water. The spatial filling patterns were designed to focus the ultrasound generated by the transducer onto different target locations on the thermochromic film. The transducer was driven at an input power of 1 W, and each location was heated for 15 s, producing a clearly visible hot spot in the thermal images captured by a thermal camera (FSFTOOLS, HP96), as shown in Supplementary Video 5. The positions of individual heating points were designed to trace a "W"-shaped pattern, as shown in Fig. 4b.

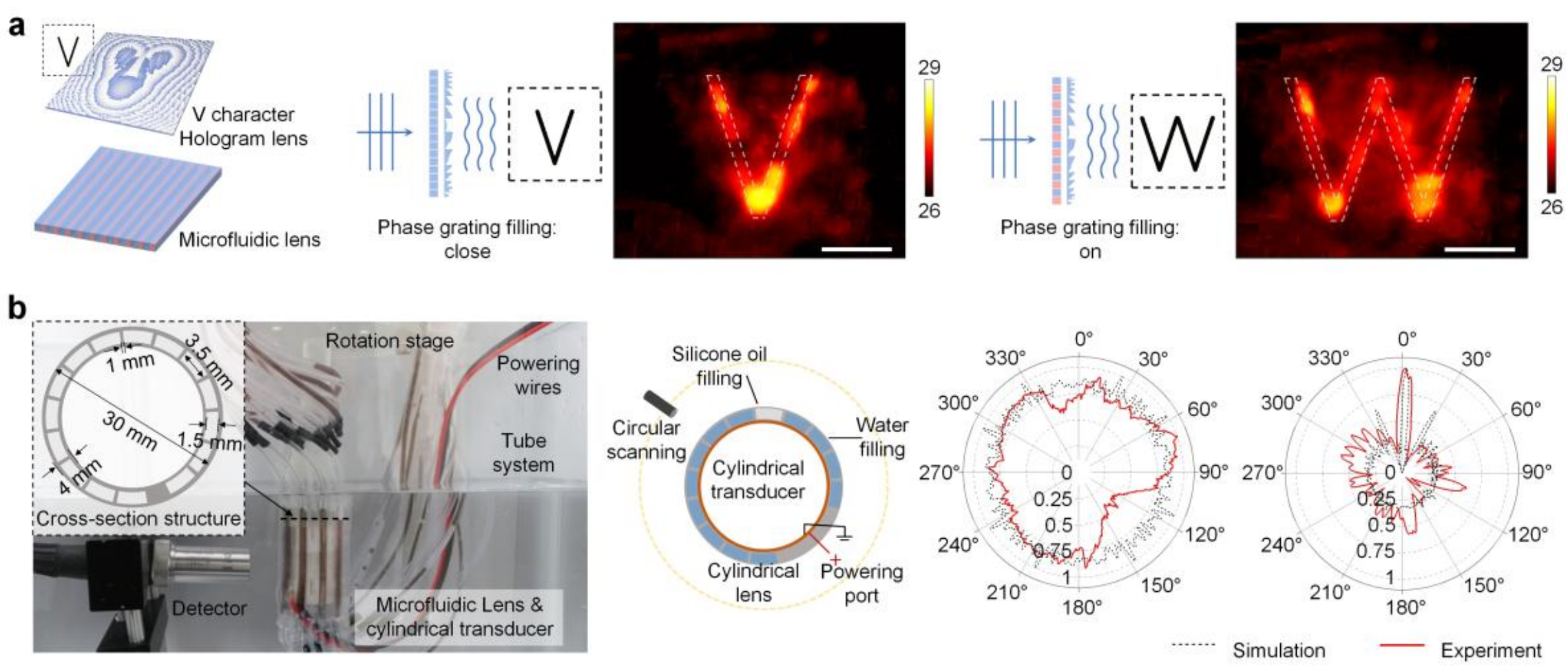

**Fig. 5 | Additional ultrasound shaping capabilities of the microfluidic lens. a.** Dynamic holographic field switching by combining a single-layer microfluidic lens with a static hologram lens. A liquid filling pattern of "101010101010101010" transforms the projected focal pattern

from a “V” shape (middle) to a “W” shape (right). The images show temperature changes on a thermochromic film induced by the resulting ultrasound field, measured using an infrared thermal camera. Scale bars, 20 mm. **b.** Dynamic modulation of polar ultrasound emission using a cylindrical microfluidic lens integrated with a cylindrical piezoelectric transducer. From left to right: schematic of the cylindrical lens architecture; experimental setup; visualization of the liquid filling pattern; and measured acoustic intensity distributions without and with the filled microfluidic lens at a distance of 10 mm from the outer surface of the lens.

## Flexible designs of microfluidic lenses

We conducted additional experiments to further demonstrate the flexibility and versatility of the reconfigurable microfluidic system. The microfluidic lens can dynamically modify the spatial field projection patterns generated by a fixed hologram, combining the high spatial and phase resolution of a static hologram lens with the reconfigurability provided by the microfluidic lens. To illustrate this capability, we recorded an infrared thermal video using a setup similar to Fig. 4a, showing how periodic liquid filling in a single-layer 1D microfluidic lens transforms the ultrasound heating pattern produced by a hologram lens from a “V” shape to a “W” shape (Supplementary Video 6). As shown in Fig. 5a, a hologram lens designed to generate a “V”-shaped focusing pattern is placed above a single layer 1D microfluidic lens. The hologram lens, fabricated by a 3D printer (Formlabs), has a spatial resolution of 50 μm and a phase resolution of 0.21 rad (derived from the depth resolution of the 3D printer). When the microfluidic lens is filled with the pattern “00000000000000,” the heating pattern on the thermochromic film forms a “V”. When the filling pattern switches to “101010101010101010”, the heating pattern correspondingly becomes a “W”.

The microfluidic lens can also be fabricated in non-planar geometries. As shown in Fig. 5b, we demonstrate a cylindrical microfluidic lens that spatially modulates the wavefront emitted by a cylindrical piezoelectric transducer, enabling dynamic control of the polar ultrasound field distribution. The microfluidic lens is placed around the cylindrical transducer. The ultrasound field is characterized by scanning along a circular trajectory around the transducer (Supplementary Note 7) at a distance of 10 mm from the outer surface of the microfluidic lens. When all channels are filled with water, the emitted ultrasound field is approximately uniform in all directions. In contrast, selectively filling a single channel with silicone oil while maintaining water in the remaining channels concentrates acoustic energy toward the angular direction corresponding to the oil-filled channel.

## Summary and Discussion

In summary, we demonstrate a framework for reconfigurable microfluidic ultrasound lenses with several key advantages over existing ultrasound modulation technologies, addressing requirements shared by both biomedical and non-biomedical ultrasound systems.

The system enables dynamic reconfigurability using a single-element transducer, eliminating the need for large multi-channel phased arrays. By selectively redistributing liquids within a lens containing microfluidic channels, an unfocused single-element source is converted into a programmable acoustic emitter capable of generating diverse spatial field patterns. This approach substantially reduces system complexity, as the number of actively controlled elements (i.e., micropumps) scales with the square root of the number of phase pixels rather than linearly. As demonstrated by a 400-pixel lens driven by only 80 independently controlled micropumps, this favorable scaling enables high spatial resolution without proportional increases in electronics, cost, or power consumption.

At the same time, the microfluidic lens offers high acoustic transmission efficiency, mechanical compliance, and thermal robustness. Binary phase modulation implemented using two liquids with similar acoustic impedance embedded in a soft matrix preserves ultrasound amplitude while enabling effective wavefront control, minimizing energy losses commonly associated with electronic or mechanically actuated modulation schemes. The soft, deformable architecture allows conformal coupling to curved or irregular surfaces, while physical separation of the lens from pumps and control electronics reduces acoustic scattering, impedance-matching constraints, and heat generation during sustained high-power operation.

The platform further offers broad frequency and material scalability. The operating frequency is determined primarily by the characteristic dimensions of the microfluidic channels and can be adapted across a wide range of ultrasound frequencies using standard microfabrication techniques. In addition to the liquids used here to demonstrate binary phase modulation, the platform is compatible with a wide variety of fluids with distinct acoustic properties, enabling flexible phase design and extension to more complex field structures.

Finally, the presented microfluidic ultrasound lens is broadly applicable across a wide spectrum of biomedical and non-biomedical ultrasound applications that share common requirements for reconfigurable and spatially programmable acoustic fields. In biomedicine, the platform is well suited for therapeutic modalities such as thermal ablation, histotripsy, lithotripsy, and neuromodulation, where acoustic energy is delivered over milliseconds to seconds and focal patterns are adjusted intermittently in response to treatment progression, tissue response, or target repositioning. The conformal, soft architecture further supports use in transcranial ultrasound, catheter-based or endoscopic ultrasound systems, and other anatomically constrained or wearable configurations. Beyond biomedical contexts, the same capabilities are relevant to non-biomedical ultrasound technologies, including acoustic manipulation and trapping, ultrasonic printing, energy delivery, and robotic systems, where reconfigurable field control, compact form factor, and reduced electronic channel count are critical. Together, these shared requirements position the platform as a general solution for programmable ultrasound control across diverse application domains.

Despite these advantages, the current implementation also highlights opportunities for further improvement. The demonstrated pattern refresh time of approximately one second, determined by the choice of liquids, channel dimensions, and pump characteristics, is well matched to quasi-static applications mentioned above but can be further reduced through engineering optimizations such as lower-viscosity fluids, increased pumping rates, or shorter fluidic paths. In addition, while the present device is fabricated using multilayer soft-lithography, future adoption of alternative fabrication strategies, including monolithic fabrication, bonding-free multilayer processes, or additively manufactured enclosed microchannels, could enhance mechanical robustness, reduce leakage risk, and improve long-term reliability. Addressing these challenges will further expand the operational envelope and practical impact of microfluidic ultrasound lenses.

## Methods

Details of the fabrication procedures for the 1D, 2D row–column, cylindrical, and holographic lenses are provided in Supplementary Note 1. The system architecture, including the FPGA, driving electronics, and graphical user interface, is described in Supplementary Note 2. Supplementary Note 3 summarizes the acoustic properties of the solid and liquid materials used in the microfluidic lens system. The acoustic field simulation method (MATLAB-based angular spectrum method) and the iterative algorithm for designing liquid-filling patterns are presented in Supplementary Note 4. A parametric study examining how lens parameters (overall size, channel count, channel dimensions, and wall thickness) affect focal resolution, efficiency, and multifocal capabilities is provided in Supplementary Note 5. Supplementary Note 7 details the measurement procedure for characterizing the acoustic field generated by the microfluidic lens and provides additional experimental information.

## Data Availability

All data needed to evaluate the conclusions in the paper are presented in the paper and/or the Supplementary Information. Additional data related to this paper can be requested from the corresponding author.

## Code Availability

All technical details for implementing the simulations and analyses are provided in the manuscript and the Supplementary Information. The full version of the code is available from the corresponding author upon reasonable request.

## Acknowledgments

C.M. and D.Z. acknowledge the funding support from National Science Foundation under Grant No. 2328096. H. J. and H. S. acknowledge the funding support from the U.S. National Institute of Biomedical Imaging and Bioengineering of the U.S. National Institutes of Health under Grant R01EB019460.

## Author Contributions

D. Z. and C.M. conceived the idea. D. Z. and C.M. conducted the theory analysis and performed the simulations. H. S. and H. J. fabricated the lenses. M. F., and F. Y. designed and fabricated the FPGA-based micropump array control system. D. Z. integrated the experimental system and performed the experiments. D. Z. and C. M. analyzed the data and wrote the manuscript. All authors participated in discussions and provided input to the manuscript. C.M. supervised the work.

## Competing Interests

The authors declare the following competing interests: a patent application (Application No. 19/371863) related to the dynamic microfluidic lens system described in this work was filed in October 2025.

# Supplementary Information

# Shape Ultrasound with Dynamic Microfluidic Lenses

*Dajun Zhang[1], Hyunji Shim[1], Michael Ferolito[1], Feng Ye[1], Hongrui Jiang[1], and Chu Ma[1]**

*[1]Department of Electrical and Computer Engineering, University of Wisconsin–Madison, Madison, WI 53706, USA*

**E-mail: chu.ma@wisc.edu*

## This file includes:

## Supplementary Note 1: Microfluidic Lenses Fabrication and Pumping System Assembly

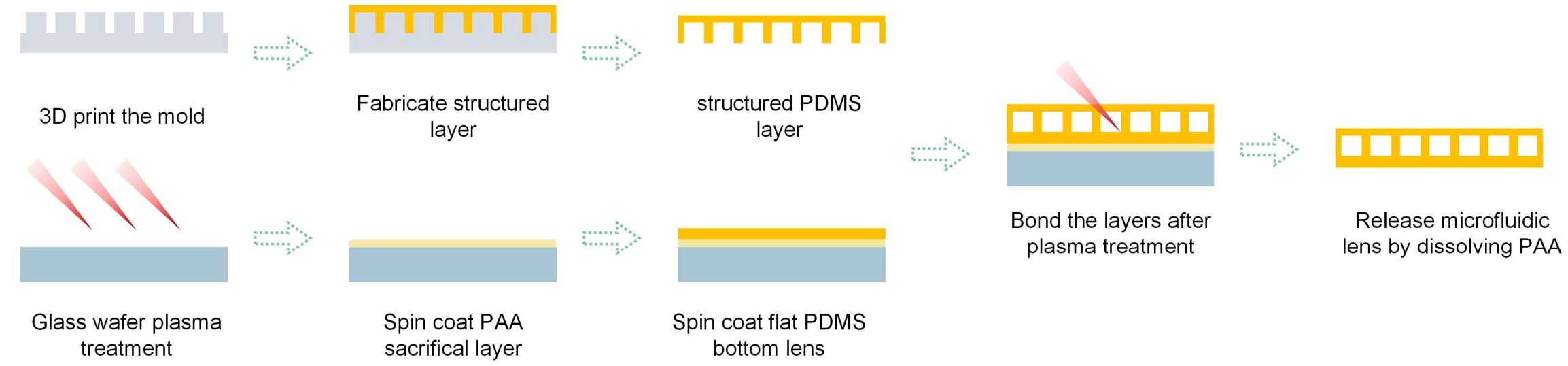

**Supplementary Fig. S1 Fabrication process of the 1D microfluidic lenses.**

The 1D microfluidic lenses were fabricated by bonding a structured PDMS layer with a planar base layer. The top structured layer was cast using a mold containing the liquid channel structure. This mold was designed via 3D modelling software and fabricated using a high-resolution 3D printer (Form 4, Formlabs). For the row-column lens arrays in this work, the mold was configured to produce 20 parallel aligned channels, each measuring 2.5 mm in width, 1.5 mm in height, and 70 mm in length, separated by 0.5 mm thick walls. At the inlet and outlet of the channels, we also designed the narrow ends to insert and fix the liquid changing tubes. The structured layer was made of polydimethylsiloxane (PDMS, Sylgard 184, Dow Corning) with a 10:1 agent ratio between base and curing. The mixture was cured at 60 °C for 30–40 min. Prior to bonding, both the structured and flat layers underwent oxygen plasma treatment (downstream Asher, YES) to mitigate potential PDMS swelling in silicone oil. Details of the procedures to overcome PDMS swelling are provided in **Supplementary Note 3**).

The bottom flat layer was fabricated separately on a glass wafer. The wafer surface was first rendered hydrophilic via plasma treatment. Subsequently, a 5 wt% poly acrylic acid (PAA) sacrificial layer was deposited on the top by spin-coated the liquid at 500 rpm for 40 s and cured at 90 °C for 1 h. A PDMS elastomer solution (15:1 ratio) was then spin-coated onto the PAA layer at 500 rpm for 40 s, yielding a uniform thickness of approximately 250 µm. After precise alignment, we placed the structured layer onto the flat layer and bonded them together with uncured PDMS as adhesive. The combined structure was then cured at 90 °C for 1 h to ensure strong adhesion, after which it was released them from the glass wafer by dissolving the PAA sacrificial layer in deionized water at 90 °C for 10 min. Finally, the fabricated lenses were baked at 130 °C overnight to stabilize the polymer network and suppress subsequent swelling of PDMS in silicone oil.

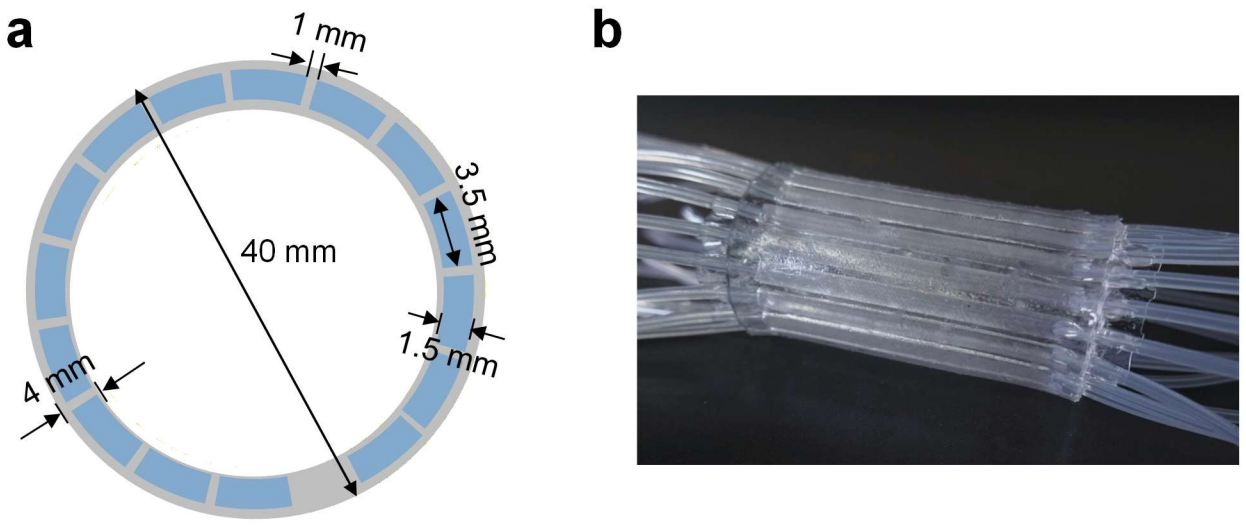

**Supplementary Fig. S2 Illustration of cylindrical lenses. a.** Structure schematic. **b.** A picture of a fabricated cylindrical lens.

The two-dimensional (2D) row-column lens (**Fig. 2**) was constructed by orthogonally stacking two identical 1D lenses. Specifically, one 1D lens (row) was rotated by 90° and aligned with a second 1D lens (column), followed by PDMS-mediated bonding. The cylindrical lens (**Supplementary Fig. S2**) was formed by conforming the 1D lens to a predefined curvature and the edges were then sealed with PDMS to maintain the cylindrical shape.

As illustrated in **Supplementary Fig. S3a**, micropumps (NKP-DC-S10B, Kamoer) were used to drive liquid flow. Individual micropumps were selectively activated to fill the target microfluidic channel with the desired liquid. All lenses were connected to the liquid pumping system via silicone tubing with an inner diameter (ID) of 0.5 mm and an outer diameter (OD) of 1.5 mm. The tubing connections at the lens inlets were mechanically reinforced and hermetically sealed with PDMS elastomer, followed by baking to fully cure the PDMS. The 0.5 mm ID/1.5 mm OD tubing connected to the microfluidic lens inlets was then coupled to the micropump tubing (3 mm ID, 5 mm OD) using heat-shrink sleeves, as shown in **Supplementary Fig. S4a**.

Because each microfluidic channel within the lens requires alternating filling with two distinct liquids to achieve an acoustic phase shift, Y-shaped three-way polypropylene connectors (1.59 mm ID, 2.3 mm OD) were employed to merge the outputs from two individual micropumps into a single lens channel. These connectors were installed within the 3 mm OD tubing segments. By selectively activating one of the two micropumps, the corresponding liquid was introduced into the channel, thereby modulating the acoustic response (see **Supplementary Fig. S4b** and **Video 3**).

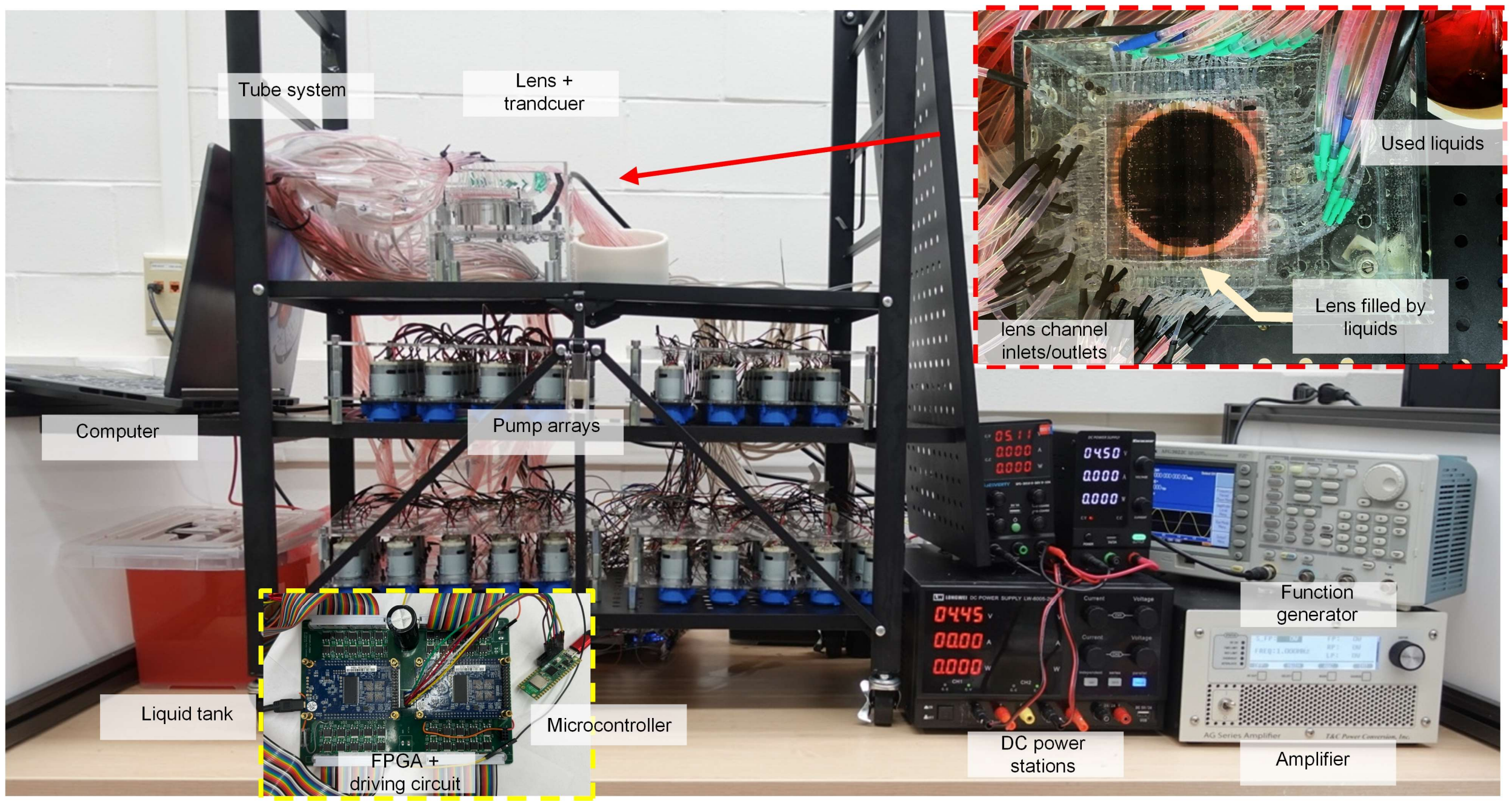

**Supplementary Fig. S3 Photograph of the microfluidic ultrasound shaping system.** The inset at the bottom (yellow dashed box) shows the FPGA-based control electronics. The inset at the top right (red dashed box) presents a top-view image of the microfluidic lens integrated with the ultrasonic transducer.

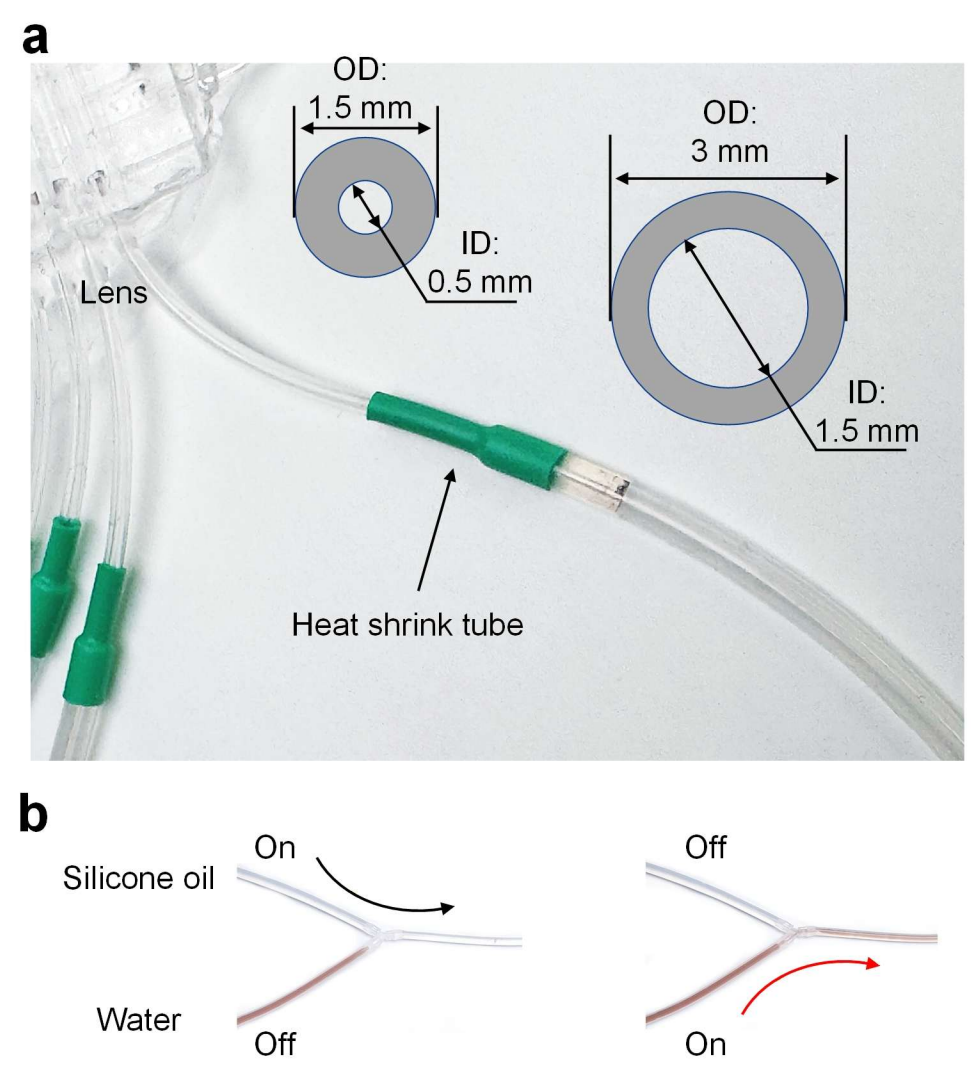

**Supplementary Fig. S4 Liquid channel connections. a.** Liquid flow at the Y-shape connector when alternating liquid selections through micropump on/off switch. **b.** Illustration of the multi-size tubing connection interface.

## Supplementary Note 2: FPGA-based micropump control and liquid changes

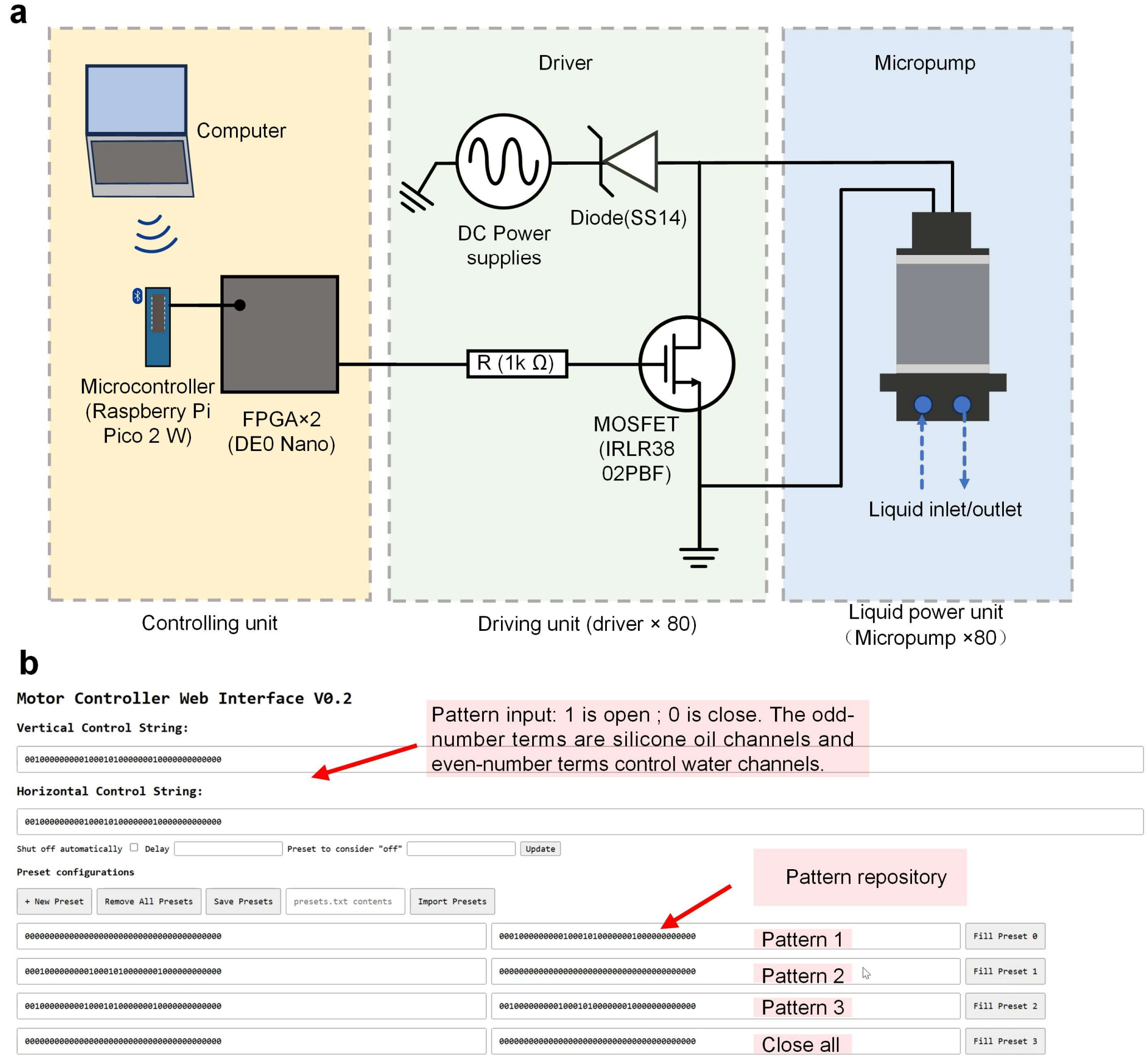

**Supplementary Fig. S5 FPGA-based micropump control system and user interface. a.** Block-level architecture of the control system with a simplified circuit diagram. **b.** Graphical user interface (GUI) illustrating three representative filling patterns stored in the pattern repository.

To achieve high-speed parallel control of 80 micropumps, FPGA chips were used to process and distribute the control signals. The system architecture and simplified circuit diagram of the micropump control module are shown in **Supplementary Fig. S5a**, which enables electrical pattern switching with response times on the order of microseconds.

Liquid filling patterns corresponding to target ultrasound field distributions are first calculated and then imported into a web-based graphical user interface (GUI) (**Supplementary Fig. S5b**), which runs on a microcontroller (Raspberry Pi Pico). The GUI encodes each liquid-filling configuration as a binary pattern string representing the on/off states of the pumps. As illustrated in **Supplementary Fig. S5b**, each row contains two 40-bit strings that together define the states of all 80 pumps in the array. These binary strings are decoded by the microcontroller and transmitted via a modified SPI interface to two FPGA broads (DE0-Nano), which wait for an external trigger pulse before updating all outputs simultaneously to ensure synchronous switching.

Each FPGA controls 40 driver circuits, which are connected directly to the micropumps. The driver circuit is a standard low-side switching configuration incorporating protection components. A 1 kΩ series resistor is placed between each FPGA output pin and the MOSFET gate to limit current flow to and from the FPGA. A flyback diode is connected in parallel with each pump to suppress voltage spikes arising from the inductive load. When the MOSFET is switched on, the output node is pulled low and the diode remains reverse-biased. Upon switch-off, the flyback diode conducts transient currents, protecting the transistor from voltage surges.

The complete system provides parallel control of 80 micropumps operating over a voltage range of 3.3 to 12 V, with individual pump currents ranging from 1 to 250 mA during operation. To supply sufficient and stable power, three programmable power supplies (SPS-3010, Jesverty) are connected in parallel. In this architecture, the FPGA controls pump switching, while the supply voltage is set independently by the power sources. Higher drive voltages correspond to faster liquid exchange within the microfluidic channels; however, to protect the microfluidic lens, the maximum operating voltage is limited to 5 V. Under these conditions, full liquid pattern reconfiguration across the lens is achieved in approximately 1 s.

## Supplementary Note 3: Material properties and sound speed measurement procedure

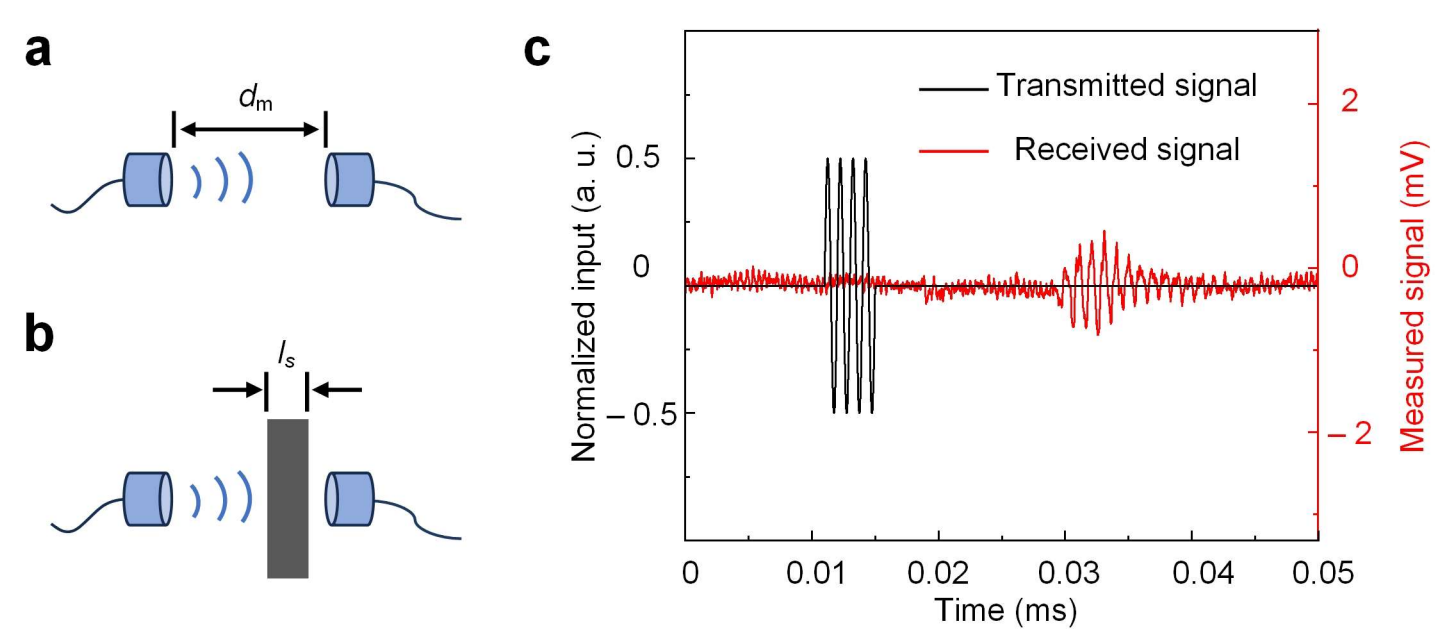

**Supplementary Fig. S6 Schematic of the sound-speed measurement setups. a. and b.** Setups for liquid (a) and solid materials (b). **c.** Representative transmitted and received signals showing a time delay.

To achieve precise phase control across the microfluidic channels, the acoustic speed of the working fluids (silicone oil and deionized water) and the structural material (PDMS) were measured at 1 MHz. The experimental setup for these measurements is detailed in **Supplementary Fig. S6**.

Two ultrasonic transducers (V303-SU, Olympus Corp.) were immersed in liquids in a face-to-face configuration, acting as a transmitter and receiver, respectively. A pulsed signal with a center frequency of 1 MHz was generated by a function generator, and the received waveforms were captured via a digital oscilloscope for subsequent analysis in MATLAB. As shown in **Supplementary Fig. S6c**, there exists a time delay $t_d$ between the transmitted and received signals. For a given liquid, two time delays were measured at different separation distances $d_1$ and $d_2$. The acoustic speed of the liquid, $c_{\text{liquid}}$, was then calculated using: $c_{\text{liquid}}=\frac{d_2-d_1}{t_{d2}-t_{d1}}$. To calculate the acoustic speed of PDMS, a sample with thickness of $l_s$ was fabricated. The time delays with ($t_{dw}$) and without ($t_{d0}$) the sample were measured, and the acoustic speed of PDMS, $c_{\text{PDMS}}$, was calculated from: $c_{\text{sample}}=\frac{l_s}{l_s/c_{water}-(t_{d0}-t_{dw})}$.

Based on the measured acoustic properties, the phase difference after the acoustic wave transmits through the channels filled with different liquids can be calculated as: $\emptyset=\left(\frac{1}{c_2}-\frac{1}{c_1}\right)2\pi fh$, where $f$=1 MHz is the operating frequency, $h$ is the channel height, and $c_1$ and $c_2$ are the acoustic speed of the two liquids. For a target phase difference $\emptyset$, the required channel height $h$ can be calculated from the same equation. Acoustic impedance, $Z$, is calculated by $Z = \rho c$. where $\rho$ is the density and $c$ is the acoustic speed. The reflection coefficient $R$ is calculated by: $R = \frac{\rho_2 c_2 - \rho_1 c_1}{\rho_2 c_2 + \rho_1 c_1}$, where $\rho_1$ and $c_1$ correspond to the transmission medium and $\rho_2$, $c_2$ to the reflecting medium.

The measured acoustic properties are summarized in **Supplementary Table S1**. Notably, the acoustic impedances of water, silicone oil, and PDMS are closely matched, minimizing undesirable reflections at the liquid-lens interfaces. This system also allows alternative materials to be incorporated as needed, enabling finer phase modulation steps beyond 0 and π.

**Supplementary Table S1. Acoustic properties of materials**

| Material | Speed of sound [m·s$^{-1}$] | Density [kg·m$^{-3}$] | Acoustic impedance [Mpa·s·m$^{-1}$] |
|---|---|---|---|
| PDMS | 1007.6 | 1015.7 | 1.023 |
| Deionized water | 1482 | 997 | 1.478 |
| Silicone oil (1cst) | 980.49 | 828.3 | 0.812 |
| Silicone oil (5 cst) | 982.32 | 915.3 | 0.899 |
| Silicone oil (1000 cst) | 987.71 | 963.9 | 0.952 |
| Alcohol (99% Isopropyl) | 1160 | 789 | 0.915 |
| Salt water | >1500[1] | >1000[1] | >1.5 |
| FC40 (Fluorinated Liquids)[2] | 640 | 1850 | 1.184 |

In this work, water and silicone oil were selected as the filling liquids. Their large difference in acoustic speed enables a thinner lens design. Both liquids are also widely used and non-toxic. Silicone oil, in particular, meets a range of medical and mechanical requirements and offers tunable viscosity depending on formulation. We additionally measured the sound speed and density of silicone oils with different viscosities and found these properties to remain stable across formulations.

However, a challenge in PDMS-based microfluidics is the absorption of low-molecular-weight species, leading to material swelling when exposed to silicone oil. The swelling speed is highly dependent on the viscosity, causing a contradiction between the liquid switching speed and lens swelling[3]. There are methods help to reduce the swelling effect[4,5]. To ensure the structural integrity of the microfluidic lenses, we systematically evaluated the degree of swelling for the lenses fabricated by different method by measuring the delamination time, which is the duration required for the planar base layer to detach from the structured channel layer under continuous exposure. A longer delamination time indicates superior resistance to swelling and enhanced interfacial adhesion.

We first investigated the correlation between silicone oil viscosity and PDMS swelling. The used PDMS samples (10:1 base-to-curing-agent ratio) were cured at 90 °C for 1 h, followed by filling the channels with silicone oils of 1 cSt and 100 cSt. In the case of 1 cSt oil, delamination occurred after 7 min, whereas no delamination was observed for 100 cSt oil even after 24 h. The same experiment was repeated with PDMS cured at 130 °C overnight. Under these conditions, delamination occurred after 20 min for 1 cSt oil, while no delamination was observed for 100 cSt oil after 24 h. These results indicate that higher-viscosity silicone oil significantly reduces PDMS swelling. However, considering the switching speed required for filling the liquid channels, 1 cSt silicone oil was selected for subsequent experiments.

Next, the effects of various PDMS treatments on the swelling behavior were systematically investigated. The treatment conditions included high-temperature curing, variation in PDMS mixing ratio, plasma treatment, and silane coating. Using the same experimental procedure, the delamination time was measured for each condition, and the results are summarized in **Supplementary Table S2**. Among the tested conditions, the combination of plasma treatment and high-temperature curing at 130 °C overnight was found to be the most effective in suppressing the PDMS swelling. However, when both the swelling resistance and the interlayer adhesion were considered, a 15:1 PDMS ratio was selected for the flat layer, while a 10:1 ratio was used for the channel layer.

**Supplementary Table S2. Comparisons of the swelling effect across different PDMS treatments**

| PDMS ratio (base: curing agent) | PDMS treatments | Delamination time |
|---|---|---|
| 10:1 | 90°C for 1 h | 7 min |
| | 130°C for overnight | 20 min |
| | Plasma treatment | 15 min |
| | Silane treatment | 10 min |
| | Plasma treatment, Silane treatment | 3 min |

|  | Plasma treatment, 130°C for overnight | 6 h |
| --- | --- | --- |
| 5:1 | 90°C for 1 h | 25 min |
|  | Plasma treatment, 130°C for overnight | >24 h |

## Supplementary Note 4: Liquid filling pattern design and field simulation

### 4.1. Iterative optimization algorithm for liquid filling patterns of the lenses

We developed an iteration algorithm based on the angular spectrum method[6] to calculate the microfluidic lens filling patterns. Assume $z = 0$ is the lens plane, and $p_0(x, y, 0)$ is the acoustic pressure after the incident wave transmits through the microfluidic lens at $z = 0$. We first separated $x$- and $y$- dependence to calculated column lens filling pattern and row lens filling pattern, respectively. When a target acoustic focal spot is located at $(x_0, y_0, z_0)$ position, we set $p_{xz}(x, z_0) = \delta(x - x_0)$ and $p_{yz}(y, z_0) = \delta(y - y_0)$ as the target acoustic pressure field $p_{xz}(x, z_0)$ and $p_{yz}(y, z_0)$, respectively.

The calculation of column lens filling pattern, denoted by $p_0(x, 0)$, are shown here as an example.

We define the field distribution at the target location as $p(x, z_0) = |p_z(x, z_0)|e^{j\angle p_z(x,z_0)}$. The angular spectrum of $p(x, z_0)$ can be expressed by Fourier transform as

$$p(x, z_0) = \frac{1}{2\pi} \int_{-\infty}^{+\infty} p(k_x, z_0) e^{j(k_x x)} dk_x$$

Similarly, the angular spectrum of the field at the lens location, $p(x, 0) = |p(x, 0)|e^{j\angle p(x,0)}$, can also be expressed as:

$$p(x, 0) = \frac{1}{2\pi} \int_{-\infty}^{+\infty} P(k_x, 0) e^{jk_x x} dk_x$$

For each $k_x$, the Fourier component at the image ($z = z_0$) and the lens ($z = 0$) planes have the following relationship:

$$P(k_x, z_0) = P(k_x, 0) H(k_x, z_0)$$

where $H(k_x, z_0) = e^{jz_0\sqrt{{k_0}^2 - {k_x}^2}}$. Here ${k_0}^2$ is $(\omega/c_w)^2$, where $\omega$ is angular frequency and $c_w$ is the speed of the sound in water.

Therefore, $p_z(x, z_0)$ can be expressed as (backward propagation):

$$p(x, z_0) = \int_{-\infty}^{+\infty} P(k_x, 0) H(k_x, z_0) e^{jk_x x}\, dk_x$$

And $p(x, 0)$ can also be expressed as (forward propagation):

$$p(x,0) = \int_{-\infty}^{+\infty} P(k_x, z_0) H(k_x, -z_0) e^{jk_x x}\, dk_x$$

We have developed an interactive algorithm[6] in MATLAB that simulates ultrasound propagation and interference in space and predicts the required lens phase distribution. The algorithm has the following steps:

a. Define the field distribution at the target locations $z$ as $p_z$. Set $p_z' = p_z$.
b. Backward propagate $p_z'$ from the target locations to the lens as $p_0{}'$
c. Convert $p_0{}'$ into binary patterns of 1 and –1 (i.e., binary phase patterns of 0 and $\pi$) as $p_{binary}{}'$ according to its phase distribution.
d. Forward propagate the binary patterns at the lens to the target locations as $p_z''$. where $k_{x,y} < \frac{2\pi}{\lambda}$ during this procedure.
e. Compare $p_z''$ and $p_z$. If their difference is smaller than a threshold (or a fixed number of iterations is reached), stop iteration. If not, go to step f.
f. Combine amplitude $|p_z|$ and phase $e^{j\angle p_z''}$ to get new $p_z'$. Jump to step b.
g. The output binary phase pattern of 0 and $\pi$ corresponds to the liquid filling pattern of the lens.

The algorithm takes the following input parameters:

- Overall size of the lens aperture.
- Number of pixels in each dimension of the lens
- Field distribution at the target image plane
- Ultrasound frequency

As the algorithm outputs the binary phase distribution at the lens plane to calculate the liquid, it also gives the simulated spatial distribution of ultrasound field.

### 4.2 Validation of MATLAB-based angular spectrum method with COMSOL full wave simulation

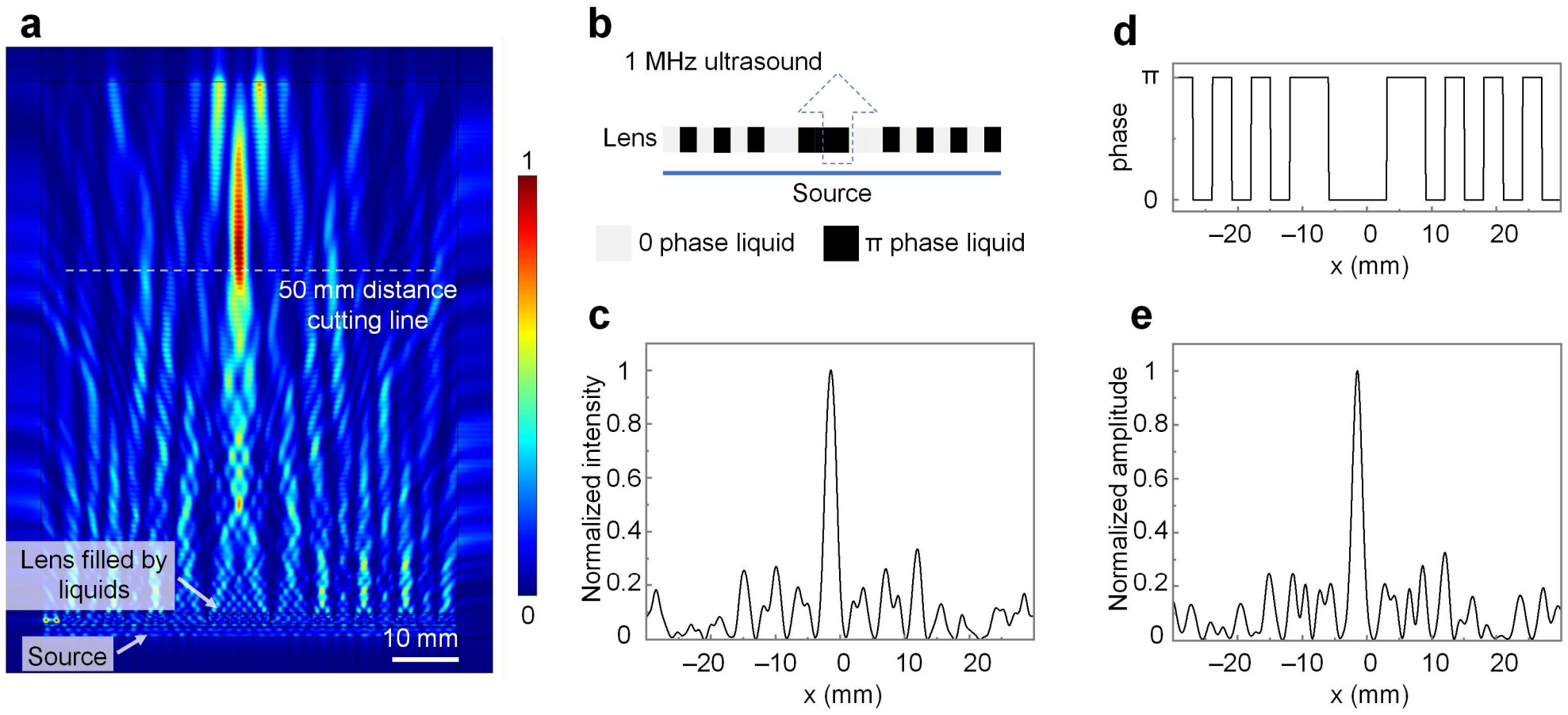

**Supplementary Fig. S7 Comparison between COMSOL Multiphysics frequency-domain simulation (Pressure Acoustics module) and the MATLAB-based angular spectrum method. a.** Normalized acoustic intensity distribution generated by a microfluidic lens with an ultrasound source positioned beneath it calculated using COMSOL Pressure Acoustics module. **b.** Liquid-filling pattern corresponding to the field distribution in (a). **c.** Normalized acoustic intensity profile extracted along the white dashed line in (a) at a propagation distance of 50 mm from the lens. **d.** Phase distribution at the lens plane used in MATLAB simulation, identical to the phase distribution generated by the liquid filling configuration in (b). **e.** Normalized acoustic intensity distribution at 50 mm from the lens calculated using the MATLAB-based angular spectrum method.

To validate the accuracy of our numerical approach, we performed a cross-validation between a full-wave finite element method simulation using COMSOL Multiphysics 6.0 and the angular spectrum method using MATLAB. The COMSOL simulation was performed in 2D in frequency domain using the Pressure Acoustics module. The 60 mm × 80 mm computational domain was enclosed by Perfectly Matched Layers (PMLs) at all boundaries to suppress unwanted reflections and simulate an infinite medium. The system consisted of a 1 MHz acoustic source and a microfluidic lens, both featuring a lateral aperture of 60 mm immersed in a water media. The ultrasound source is put close to the lens interface, transmitting a 1MHz plane wave propagating in the *z*–direction, where the ultrasound lens is filled by water and silicone oil to provide phase shift difference of 0 and π.

A representative normalized acoustic intensity distribution from the COMSOL simulation was shown in **Supplementary Fig. S7a**. The normalized acoustic filed intensity is defined as $I_{norm}(x,y) = \frac{I(x,y)}{\max(I(x,y))}$, where max $(I(x,y))$ is the maximum intensity in each plot. $I(x,y) =$

$\frac{p_{rms}{}^2}{\rho_w c_w}$, where $p_{rms}$ is the root mean square pressure $\frac{p_{peak}}{\sqrt{2}}$ from simulation, $\rho_w$ is the density of the water and $c_w$ is the acoustic speed in water. The liquid filling pattern, and normalized acoustic intensity along the cutting line at $z$ = 50 mm is shown in **Supplementary Fig. S7b** and **S7c**, respectively.

In parallel, the same phase distribution was modeled using the angular spectrum method in MATLAB, as shown in **Supplementary Fig. S7d** and **S7e**. The angular spectrum method decomposes the initial complex pressure field into its angular spectrum components and propagates them in the spatial frequency domain, offering a computationally efficient alternative to FEM for large-scale field mapping. To quantify the agreement between the two methods, the normalized acoustic intensity at the $z$ = 50 mm cutting line from two simulation methods are both shown for the comparison. As demonstrated in the results, the intensity distributions obtained from both methods exhibit excellent agreement, with nearly identical peak positions and side-lobe characteristics. This high degree of consistency validates the efficacy of the angular spectrum method for predicting the 2D acoustic field distributions presented in this study. Consequently, the MATLAB-based angular spectrum method was utilized as the primary numerical tool for the iterative design and large-scale simulation of the acoustic holograms and dynamic lenses.

## 4.3 More examples of simulated and experimentally measured field patterns

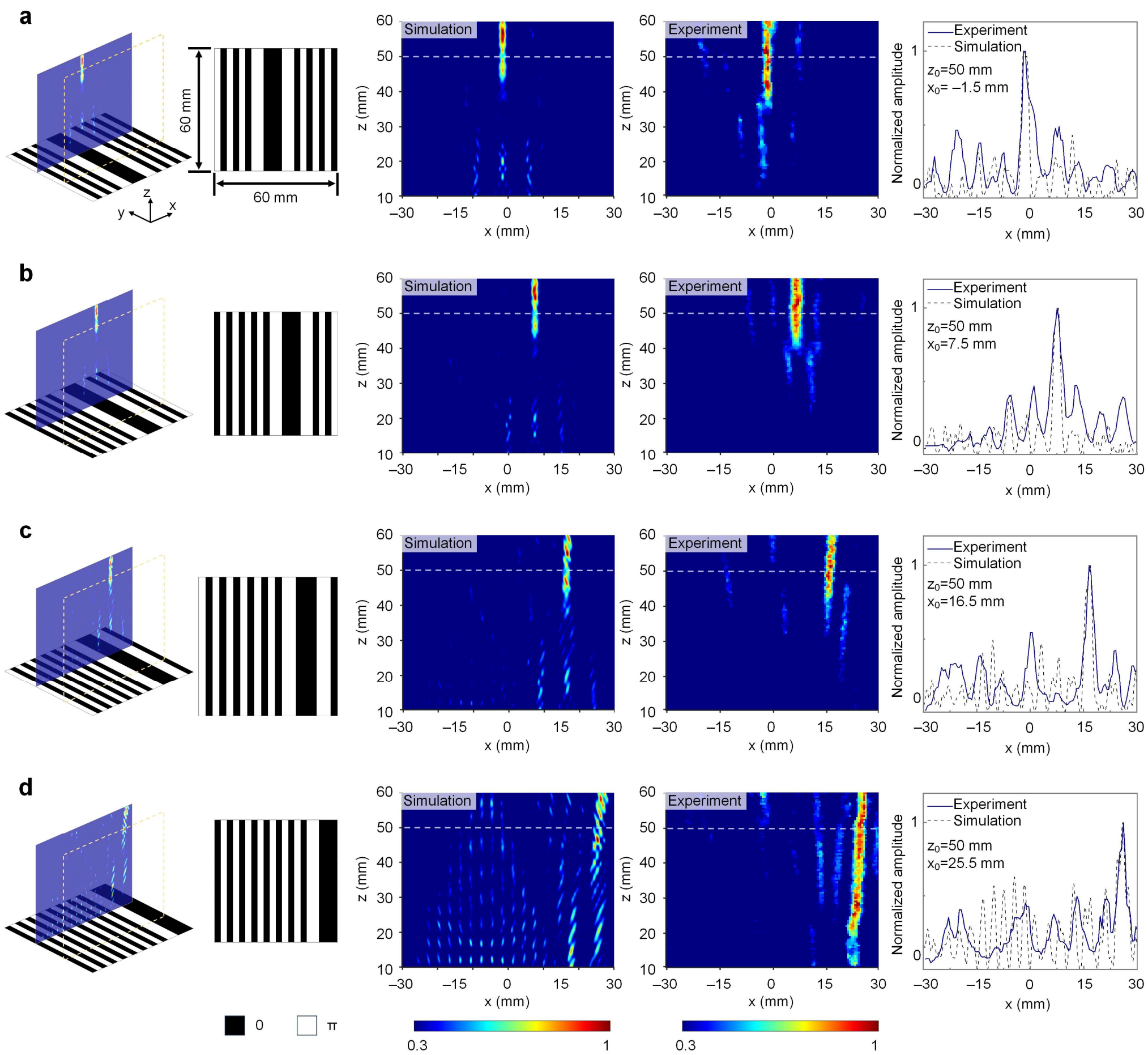

**Supplementary Fig. S8 Ultrasound focusing patterns generated by a one-dimensional microfluidic lens at varying lateral positions.** From left to right, each row shows: the simulated acoustic intensity and corresponding phase profile at the lens plane; the simulated normalized acoustic intensity distribution in the $y = 0$ plane; the experimentally measured normalized acoustic intensity distribution in the plane; and normalized acoustic intensity distribution extracted along the white dashed lines. Target focal positions are: **a.** $x_0 = -1.5$ mm, $z_0 = 50$ mm; **b.** $x_0 = 7.5$ mm, $z_0 = 50$ mm; **c.** $x_0 = 16.5$ mm, $z_0 = 50$ mm; **d.** $x_0 = 25.5$ mm, $z_0 = 50$ mm**.**

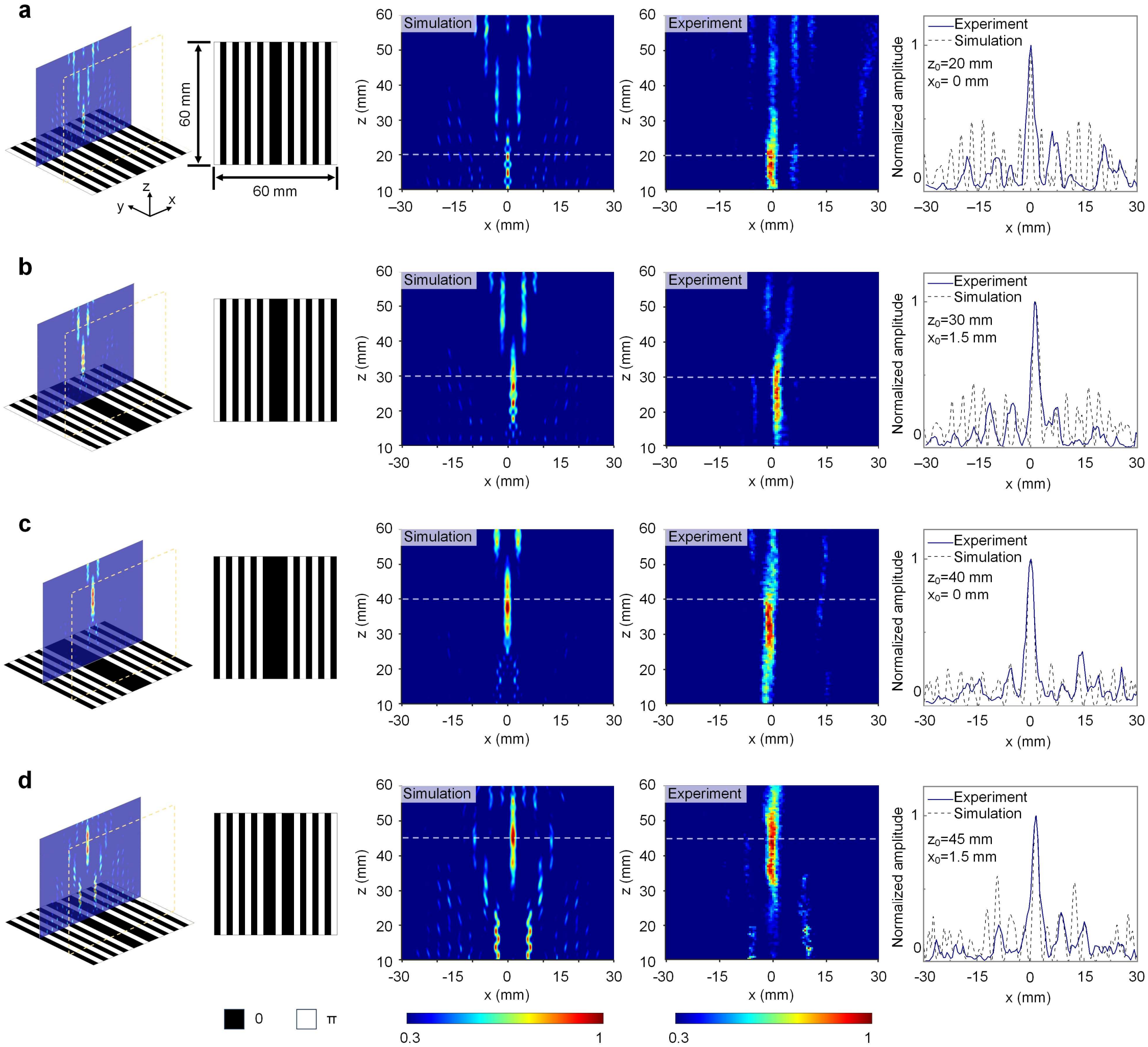

**Supplementary Fig. S9 Ultrasound focusing patterns generated by a one-dimensional microfluidic lens at varying focal distance.** From left to right, each row shows: the simulated acoustic intensity and corresponding phase profile at the lens plane; the simulated normalized acoustic intensity distribution in the $y = 0$ plane; the experimentally measured normalized acoustic intensity distribution in the $y = 0$ plane; and normalized acoustic intensity distribution extracted along the white dashed lines. Target focal positions are: **a.** $x_0 = -1.5$ mm, $z_0 = 50$ mm; **b.** $x_0 = 7.5$ mm, $z_0 = 50$ mm; **c.** $x_0 = 16.5$ mm, $z_0 = 50$ mm; **d.** $x_0 = 25.5$ mm, $z_0 = 50$ mm.

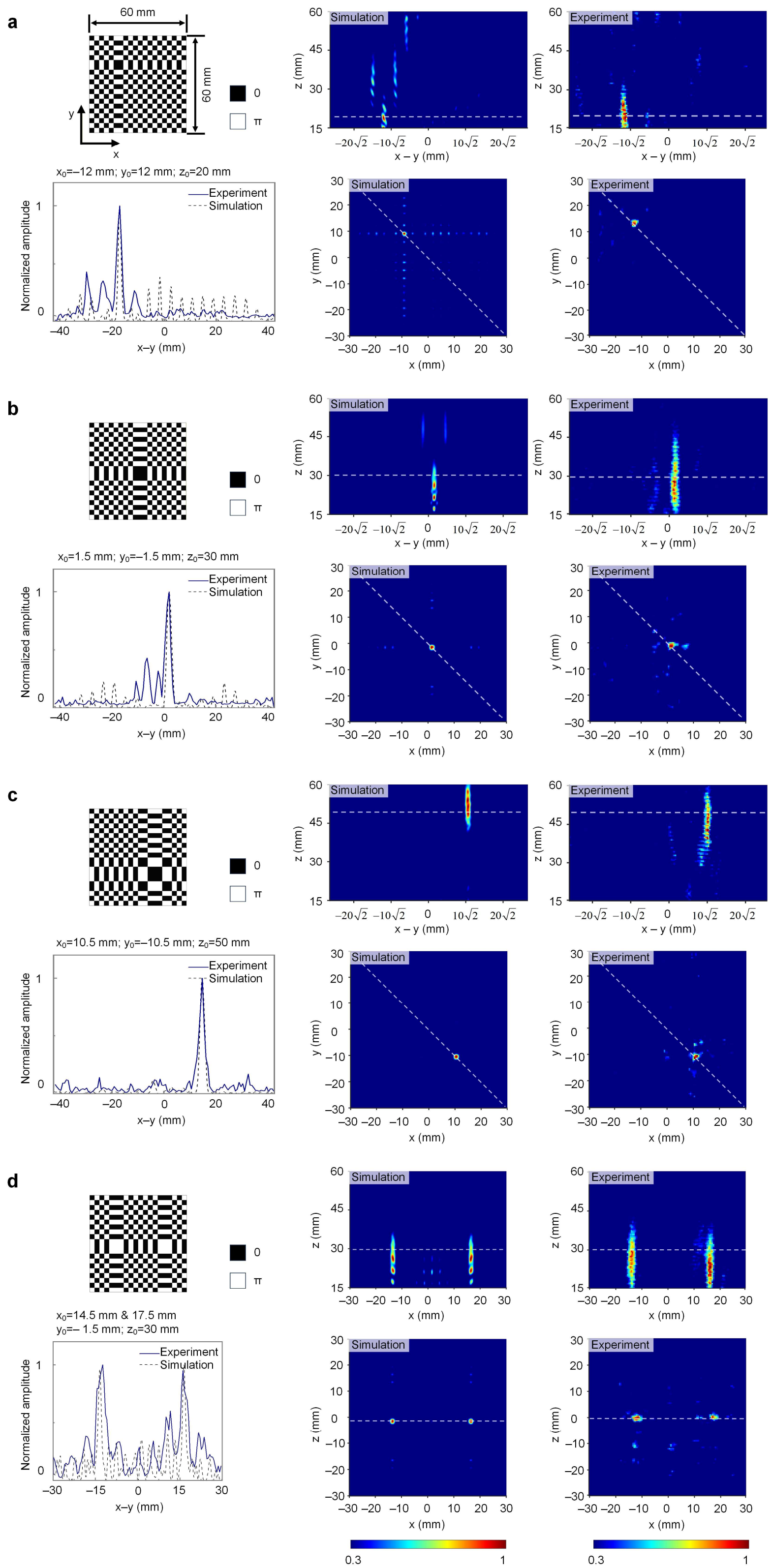

**Supplementary Fig. S10 Ultrasound focusing patterns generated by the two-dimensional microfluidic lens.** For each panel (a–d), the first row shows, from left to right: the phase profile

at the lens plane; the simulated acoustic intensity distribution in the $x$–$z$ plane ($y = 0$) for panels a–c and in the $x$–$z$ plane ($y = -1.5$ mm) for panel d; and the corresponding normalized acoustic intensity distribution measured experimentally in the same planes. The second row shows the normalized acoustic intensity distribution extracted along the white dashed lines in the two-dimensional cutting planes, followed by comparisons of simulated and experimentally measured acoustic intensity distributions in the *x*–*y* plane at the target focal distances. The target focal positions are: **a.** $x_0 = -12$ mm, $y_0 = 12$ mm, $z_0 = 30$ mm; **b.** $x_0 = -1.5$ mm, $y_0 = -1.5$ mm, $z_0 = 50$ mm; **c.** $x_0 = 10.5$ mm, $y_0 = 10.5$ mm, $z_0 = 50$ mm; **d.** $x_0 = 14.5$ mm and $-17.5$, $y_0 = -1.5$ mm, $z_0 = 30$ mm.

## Supplementary Note 5: Parametric study of the field pattern resolution and focusing efficiency

### 5.1 Theoretical limitations of focusing resolution

The focal spot size generated by the microfluidic lens cannot be infinitely small, because of several physical limits.

**a. Limit due to the pixel size ($d$)**

Because of Shannon's sampling theorem, the pixel size $d$, which is determined by the channel width, can only support limited $k_x$ at the lens plane at $z = 0$:

$$k_x < \frac{\pi}{d}$$

**b. Limit due to evanescent wave:**

A spectrum component, i.e., a plane wave, has the expression of $P_{k_x}(x,z) = P_0 e^{i(k_x x + k_z z)}$, where $k_z = \sqrt{k_0^2 - k_x^2}$. If $k_x > k_0$, $k_z$ becomes an imaginary number, $k_z = ia$.

Then,

$$P_{k_x}(x,z) = P_0 e^{i(k_x x + k_z z)} = P_0 e^{ik_x x} e^{-az}$$

The wave becomes evanescent wave whose amplitude decays exponentially with z. Therefore, this wave component cannot carry energy between the lens and the target image, leading to a resolution limitation of:

$$k_x < \frac{2\pi}{\lambda}$$

Summarizing Limits a and b, we get the maximum lateral wavenumber that can be supported by the lens, $k_{x,max} = \min\{\frac{\pi}{d}, \frac{2\pi}{\lambda}\}$. The focal spot can be approximated as $p(x) = \frac{1}{2\pi}\int_{-k_{x,max}}^{+k_{x,max}} e^{jk_x x} dk_x = \frac{\sin(k_{x,max}x)}{\pi x}$ [7]. The limited spectrum bandwidth leads to a focal spot main peak width of:

$$R = \frac{2\pi}{k_{x,max}} = \max\{d, \frac{\lambda}{2}\}$$

Here the main peak width is defined as half of the zero-crossing width of the main peak.

**c. Limit due to finite lens aperture:**

Assume the overall size of the lens aperture is $D$, the wavelength is $\lambda$, and the distance between the lens and the focal spot is $z_0$. Based on the theory of Fresnel diffraction[7], the focal spot $p(x)$ can be approximated as the Fourier transform of a rectangle function:

$$p(x) = \frac{1}{2\pi} \int_{-k_{x,max}}^{+k_{x,max}} \mathrm{rect}(\frac{x'}{D}) e^{-i\frac{2\pi}{\lambda z_0}x'} dx' = \frac{\sin\left(\pi \frac{D}{\lambda z_0} x\right)}{\frac{\pi x}{\lambda z_0}}$$

The focal spot size is

$$R = \frac{\lambda z_0}{D}$$

Which is inversely proportional to D.

Overall, because of the Limits a-c, the focal spot size is:

$$\mathrm{R} = \max\{d, \frac{\lambda}{2}, \frac{\lambda z_0}{D}\}$$

In our lens design, $d = 2\lambda$, and $D$ is 6 cm. When the focal distance $z_0$ is greater than 12 cm, the focal spot size is $\frac{\lambda z_0}{D}$ and is mainly limited by the lens aperture size $D$. When the focal distance $z_0$ is smaller than 12 cm, the focal spot size is $d$, which is the pixel size.

We also performed numerical simulations to verify these limits. **Supplementary Figure S11** presents the lens phase patterns and corresponding ultrasound field distributions for varying lens aperture sizes $D$, while keeping the pixel size fixed at $d = 3$ mm and the focal distance at $z_0 = 50$mm. Theoretically, the focal spot size is determined by $\max\left\{d, \frac{\lambda}{2}, \frac{\lambda z_0}{D}\right\}$. Under the current conditions, the pixel size $d$, which is approximately $2\lambda$, is the dominant limiting factor. Consequently, as $D$decreases from 120 mm to 24 mm in the simulations, the focal spot size, defined as the full width at half maximum (FWHM) of the main lobe, increases only slightly, from 2.0 mm to 3.2 mm.

**Supplementary Figure S12** demonstrates the influence of the pixel size $d$ on focusing performance for a fixed lens aperture size of 60 mm and focal distance of 50 mm. Under the condition where $\max\left\{d, \frac{\lambda}{2}, \frac{\lambda z_0}{D}\right\} = d$, a reduction in $d$ from 3.75 mm to 1.5 mm leads to a proportional sharpening of the focal spot. This pronounced decrease in spot size underscores the critical role of the tube width, which determines the pixel size, in our lens design.

**Supplementary Figure S13** demonstrates the influence of the focal distance $z_0$ on focusing performance for a fixed lens aperture size of 60 mm and pixel size of 3 mm. The impact of $z_0$ remains smaller compared to the geometric constraint imposed by $d$. Increasing $z_0$ from 2.5 cm to 5 cm does not change the focal spot size. And further increasing $z_0$ from 5 cm to 10 cm changes the focal spot size from 2 mm to 2.4 mm.

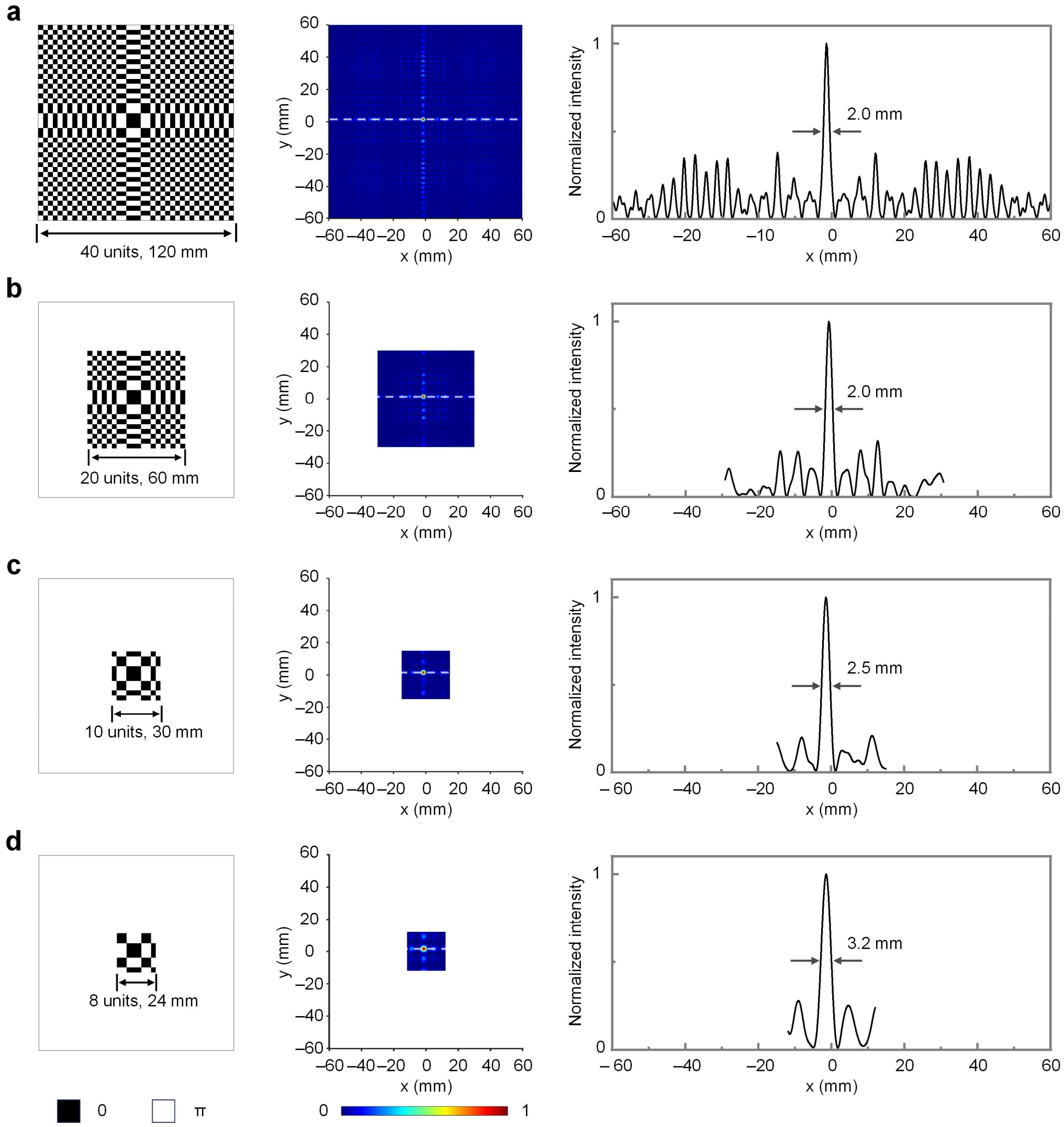

**Supplementary Fig. S11 Numerical simulations of acoustic fields for various lens apertures.** From left to right: the phase patterns at the lens plane, normalized intensity at $z = 50$ mm plane, and normalized intensity along the cutting line at peak ($z = 50$ mm, $y = 1.5$ mm; indicated by white dashed lines). The width of the tubes is all 3 mm, and focal distance is 50 mm. The size of lens aperture varies as follows: **a.** 120 mm; **b.** 60 mm; **c.** 30 mm; **d.** 24 mm.

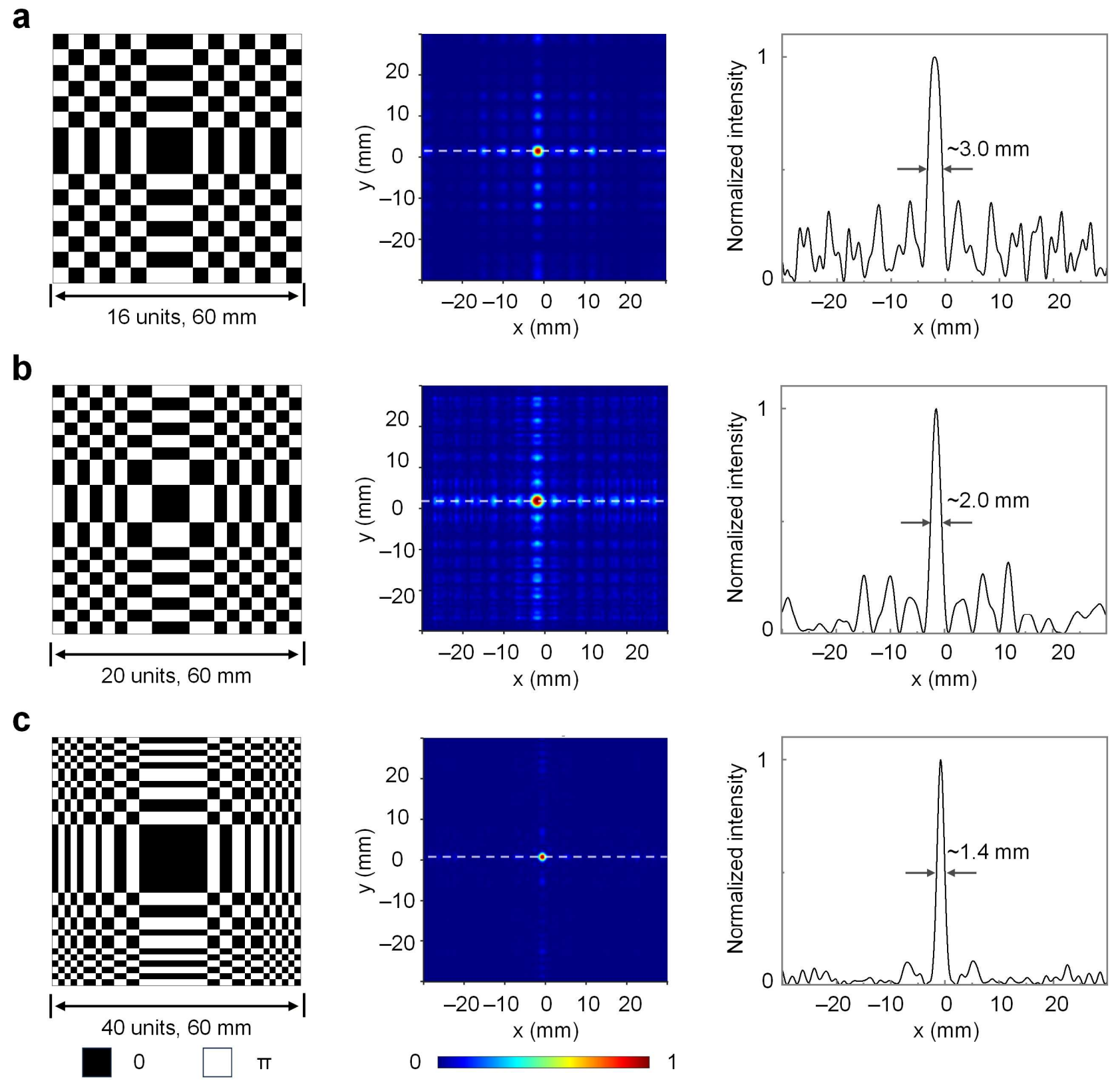

**Supplementary Fig. S12 Numerical simulation of acoustic field distributions for various pixel sizes.** From left to right: the phase patterns at the lens plane, normalized intensity at the focal plane ($z = 50$ mm), and normalized intensity along the cutting line at the peak position ($z = 50$ mm and $y = 1.5$ mm, indicated by white dash line in 2D field distribution). The results are shown for various pixel sizes: **a.** $d = 3$ mm; **b.** $d = 3.75$ mm; **c.** $d = 1.5$ mm.

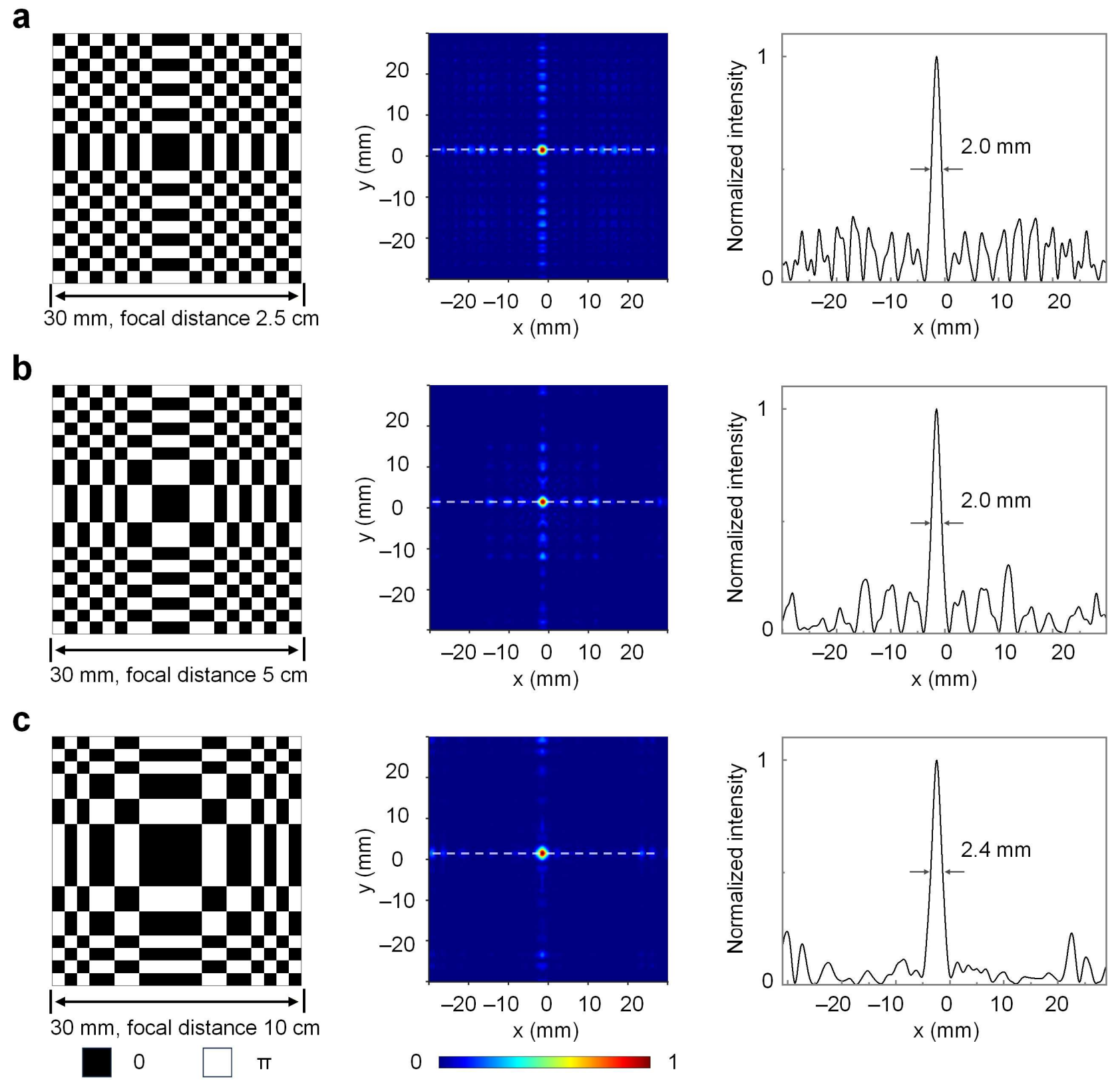

**Supplementary Fig. S13 Numerical simulation of acoustic field distributions for various focal distances.** From left to right: phase profiles at the lens plane, normalized intensity at the designated focal plane ($z = 50$ mm), and normalized intensity along at the peak position (indicated by white dashed lines in the 2D fields). The results are shown for various focal distances: **a.** 50 mm; **b.** 25 mm; **c.** 100 mm.

## 5.2 Ability to generate multiple focal spots

The row-column lens also enables the generation of multiple focal spots. **Supplementary Figure S14** illustrates various multi-focal patterns alongside their corresponding numerical simulations. The generated focal spots exhibit high uniformity in both intensity and shape. Using a lens with $20 \times 20$ pixels, we can generate $2 \times 2$, $3 \times 3$ and $4 \times 4$ focal spots. Notably, due to the inherent geometric constraints of the row-column design, the focal spots are restricted to orthogonal array configurations. And when the number of focal spots increases to $4 \times 4$, the side lobes become non-negligible, indicating the need to increase the number of pixels in the lens design.

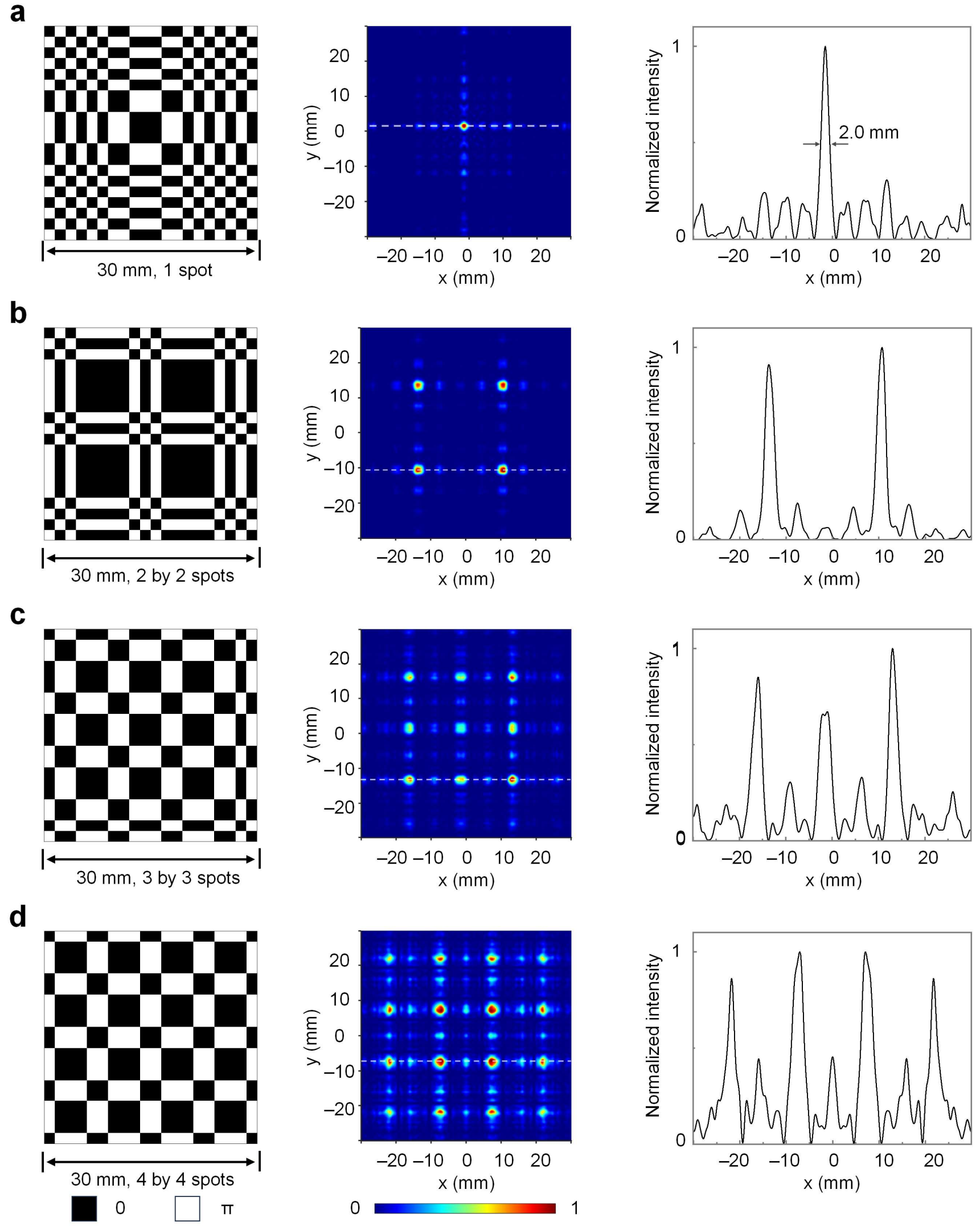

**Supplementary Fig. S14 Numerical simulations of multi-focal acoustic patterns.** From left to right: the phase patterns at the lens plane ($z = 0$), the normalized intensity at the focal plane ($z_0 = 50$ mm), and normalized intensity along the peak positions (indicated by white dash cutting line in 2D field distribution). The target positions for each pattern are as follows: **a.** single focal spot at $x = -1.5$ mm and $y = 1.5$ mm; **b.** $2 \times 2$ focal spots: $x = -13.5$ mm; $10.5$ mm and $y = -10.5$ mm; $13.5$ mm. **c.** $3 \times 3$ focal spots: $x = -16.5$ mm; $-1.5$ mm; $13.5$ mm; $16.5$ mm and $y = -16.5$ mm; $-1.5$ mm; $13.5$ mm; **d.** $4 \times 4$ focal spots: $x = -22.5$ mm; $-10.5$ mm; $1.5$ mm; $13.5$ mm; and $y = -22.5$ mm; $-10.5$ mm; $1.5$ mm; $13.5$ mm.

### 5.3 Influence of channel wall thickness and the separation distance between the row and column layers

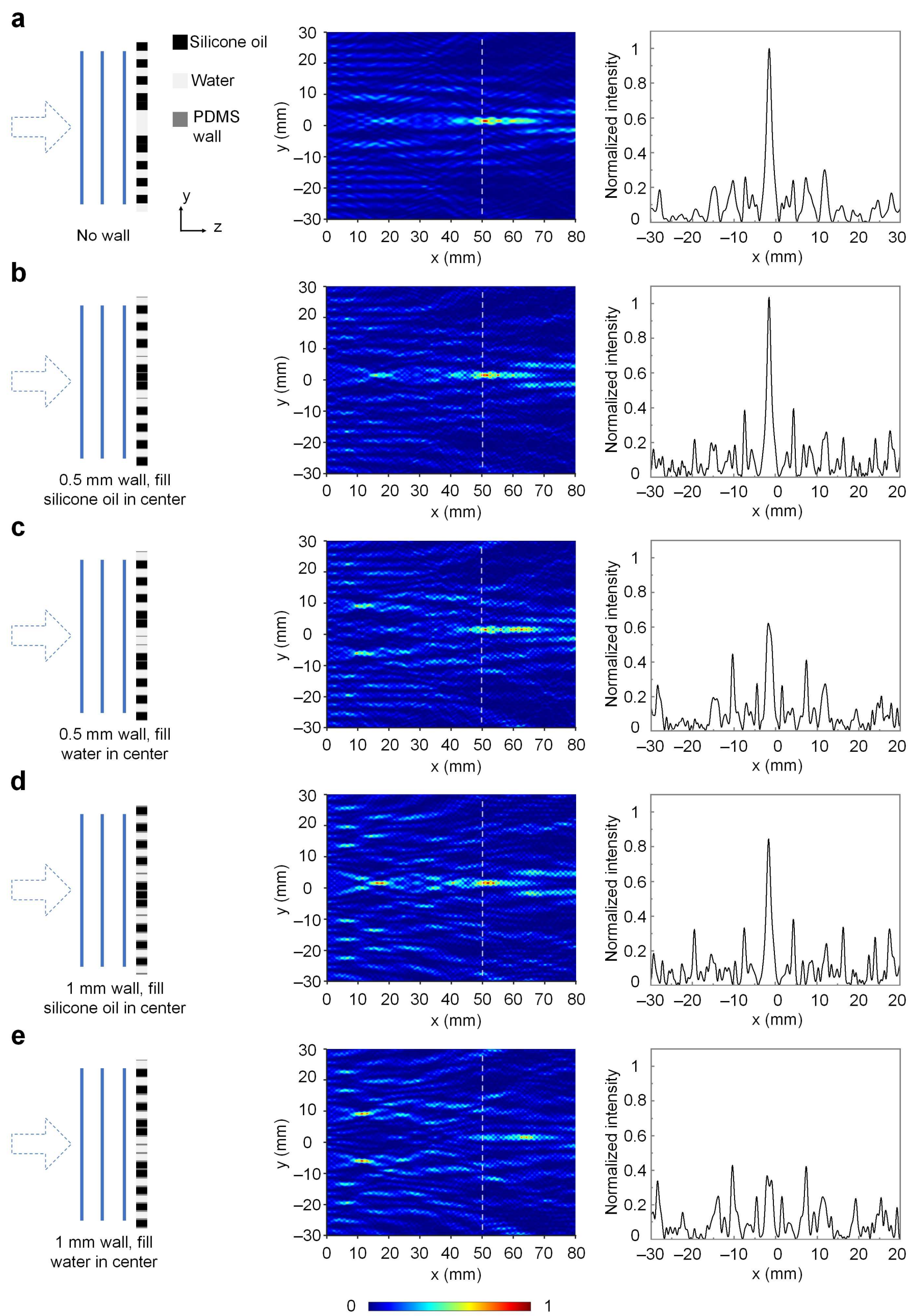

**Supplementary Fig. S15 Influence of liquid channel walls on focusing performance.** Simulations are performed for a one-dimensional (1D) row/column lens with a target focal position at $x_0 = -1.5$ mm and $z_0 = 50$ mm. From left to right, the columns show: phase profiles at the $z = 0$ plane with simulation schematics; normalized acoustic field intensity over the full 2D region; and normalized field intensity profiles extracted along the white dashed lines. **a.** Phase pattern without internal channel walls. **b.** Wall thickness of 0.5 mm, with central channels filled with silicone oil producing the same transmitted phase as the PDMS walls. **c.** Wall

thickness of 0.5 mm, with central channels filled with water, generating a π phase difference relative to the PDMS walls. **d.** Wall thickness of 1 mm, with central channels filled with silicone oil producing the same transmitted phase as the PDMS walls. **e.** Wall thickness of 1 mm, with central channels filled with water, generating a $\pi$ phase difference relative to the PDMS walls.

In practice, several structural factors can influence the focusing performance of the acoustic lens. The lens architecture requires internal walls to provide structural support and separate the microfluidic channels. These walls may provide unfavorable wave scattering because of their distinct acoustic properties with the needed infilling liquids. In this study, we characterized the acoustic properties of the wall material, polydimethylsiloxane (PDMS) (refer to **Supplementary Note 3** for details). Our results indicate that PDMS exhibits an acoustic velocity comparable to that of the silicone oils used. We subsequently simulated the impact of these internal walls on ultrasound beam shaping. As shown in **Supplementary Fig. S15a**, we initially considered a baseline pattern without walls, where the phase difference is generated solely by channels filled with disparate liquids. The target focal distance of this pattern is $z_0 = 50$ mm, characterized by a well-defined focal spot. To quantify the influence of the walls, we recorded the peak intensity of the baseline no-wall simulation at $z = 50$ mm, $I_{max}$, and utilized it to normalize the simulated field intensity for configurations containing walls: $I_{norm}(x, y) = \frac{I(x,y)}{I_{max}}$, The normalized intensity profiles, $I_{norm}(x, y)$ are presented in **Supplementary Figs. S15b–e**. The internal walls lead to a reduction in peak intensity at the focal spot. Notably, the choice of infilling liquid plays a pivotal role in maintaining performance: when the central channels are filled with silicone oil, which closely matches the acoustic impedance and speed of PDMS, the lens demonstrates significantly higher efficiency compared to configurations where the central channels are filled with water. The same ultrasound field pattern can be produced by adding a constant $\pi$ phase shift to all the pixels. Therefore, we should always choose the pattern distributions where the critical channels (e.g., the central ones) are filled with silicone oil.

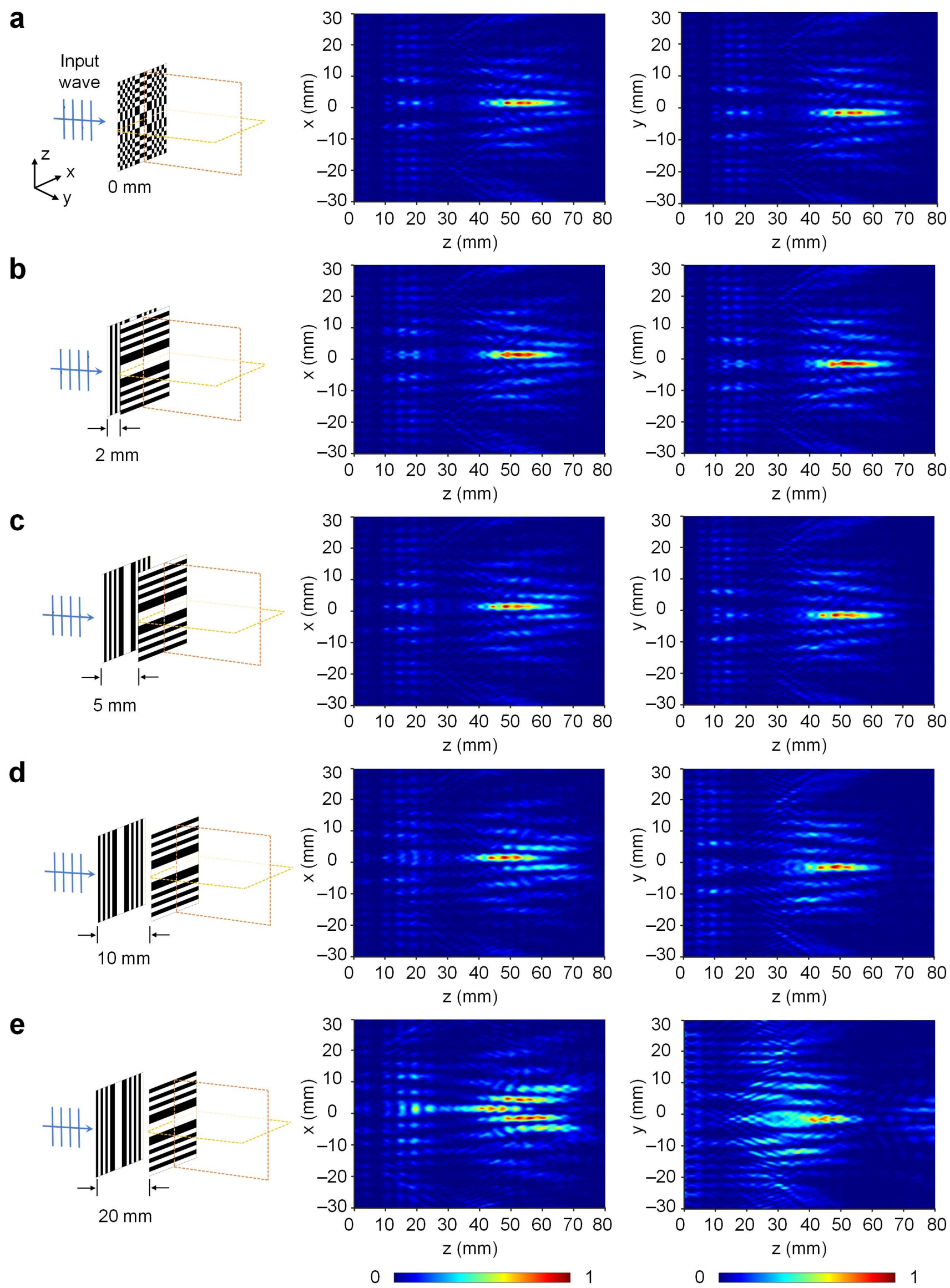

**Supplementary Fig. S16 Influence of the separation distance between the row and column layers on focusing performance.** From left to right, the columns show: schematics illustrating the row and column phase patterns and their separation distance; normalized acoustic field intensity in the $x$–$z$ plane through the focal center; and normalized acoustic field intensity in the $y$–$z$ plane through the focal center. **a.** No separation between the row and column layers. **b.** Separation distance of 2 mm. **c.** Separation distance of 5 mm. **d.** Separation distance of 10 mm. **e.** Separation distance of 20 mm.

Another practical consideration is the separation distance between the row and column layers, which is 2 mm in the current lens implementation. We conducted simulations over a range of separation distances to evaluate their impact on focusing performance, as summarized in **Supplementary Fig. S16**. The results show that the focusing effects of the row and column layers are additive, with each layer independently controlling focusing along one spatial dimension. However, introducing a gap between the two layers inevitably shifts the focal position along the $z$-axis. When the separation becomes excessively large, such as 10 mm or 20 mm, as shown in **Supplementary Figs. S16d** and **e**, the focal position shifts forward or backward, leading to degraded focusing performance and redistribution of acoustic energy toward off-axis or side-lobe beams.

## Supplementary Note 6: More results on combining microfluidic lens with acoustic hologram

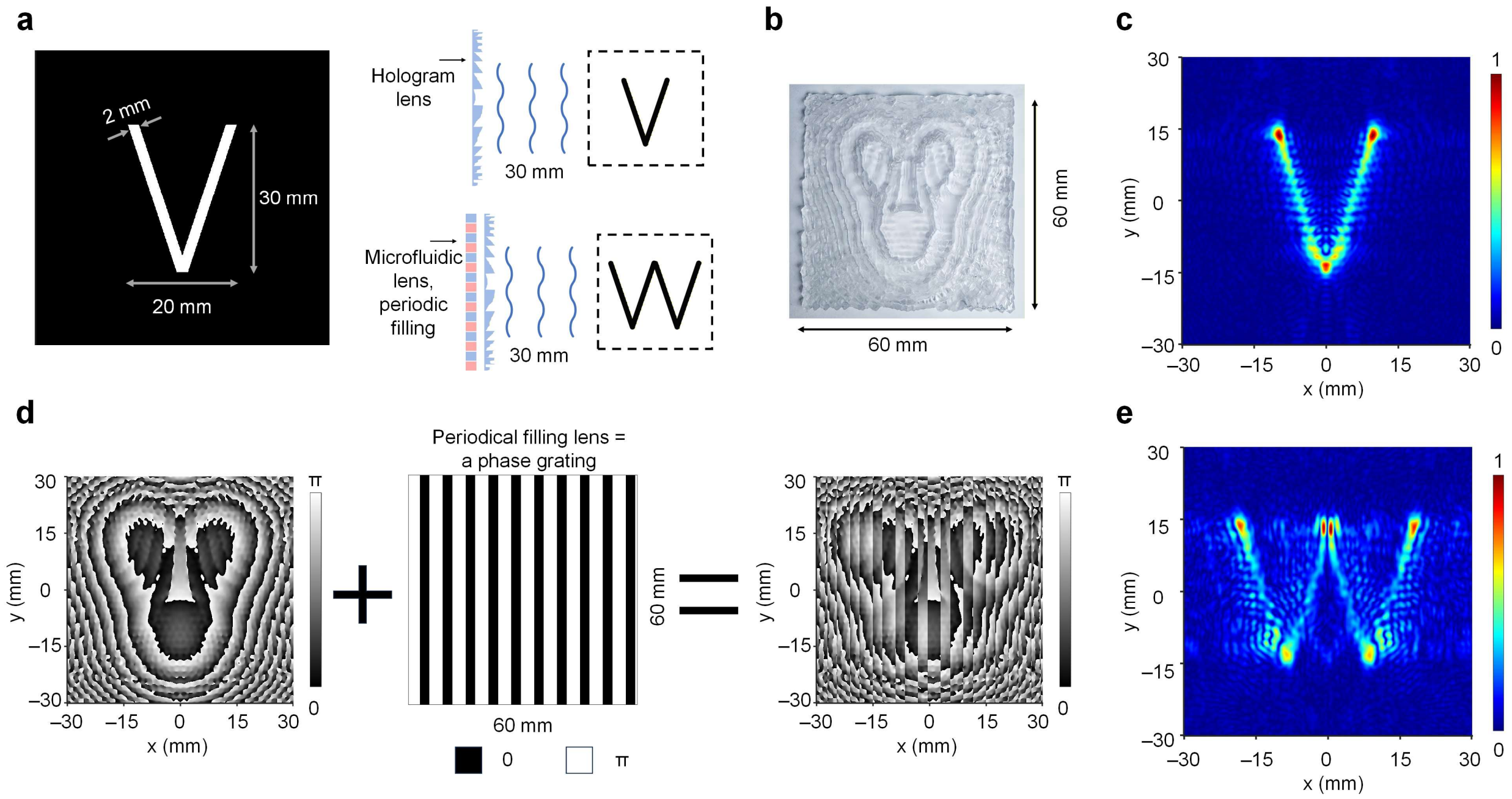

**Supplementary Fig. S17 Dynamic reconfiguration of an acoustic hologram from a V shape to a W shape using a 1D microfluidic lens as a phase grating. a.** Target focal plane field pattern and schematic illustration of the experimental setup. **b.** A photo of the fabricated "V"-shaped hologram lens. **c.** Simulated ultrasound field intensity at the focal plane ($z = 30$ mm) corresponding to the "V"-shaped hologram. **d.** Phase modulation introduced by applying a phase grating with a period of 3 mm. **e.** Simulated ultrasound field intensity at the focal plane after superposition of the hologram phase and the grating phase, resulting in a "W"-shaped pattern.

Integrating the microfluidic lens with an acoustic hologram enables dynamic, high-spatial-resolution control of ultrasound fields. **Supplementary Figure S17** illustrates the implementation of a dynamic phase array that morphs an acoustic hologram from a "V"-shaped pattern to a "W"-shaped pattern, corresponding to the demonstration shown in **Fig. 5** of the main text. In the experimental setup, a static holographic lens designed to generate a V-shaped focal pattern at $z = 30$ mm is combined with a single-layer 1D microfluidic lens that is used as a binary phase grating, as shown in **Supplementary Fig. S17a.** The required phase map was calculated following the procedure described in **Supplementary Note 4**, and the corresponding surface height profile of the hologram lens was determined by $h_{(x,y)} = h_0 + \frac{\varphi(x,y)}{2\pi} \times \frac{\lambda_{water}\lambda_{PDMS}}{\lambda_{water}-\lambda_{PDMS}}$. PDMS holographic lenses were fabricated using the same process as the microfluidic lenses. After designing the target acoustic hologram phase map, it was converted into a height profile, and a counter-mold was fabricated using a 3D printer (Formlabs). The counter-mold were then used to produce the “V”-shaped hologram lens.

The fabricated hologram lens and the corresponding simulated acoustic intensity at the target plane are shown in **Supplementary Figs. S17b** and **c**, respectively. By dynamically modulating the liquid filling pattern within the microfluidic layer, the cumulative phase profile at the lens plane is altered (**Supplementary Fig. S17d**). As a result, the acoustic hologram is reconfigured into a “W”-shaped field pattern at the focal plane, as demonstrated in **Supplementary Fig. S17e**.

## Supplementary Note 7: Experimental details

### 7.1 Ultrasound field characterization for the row-column lens

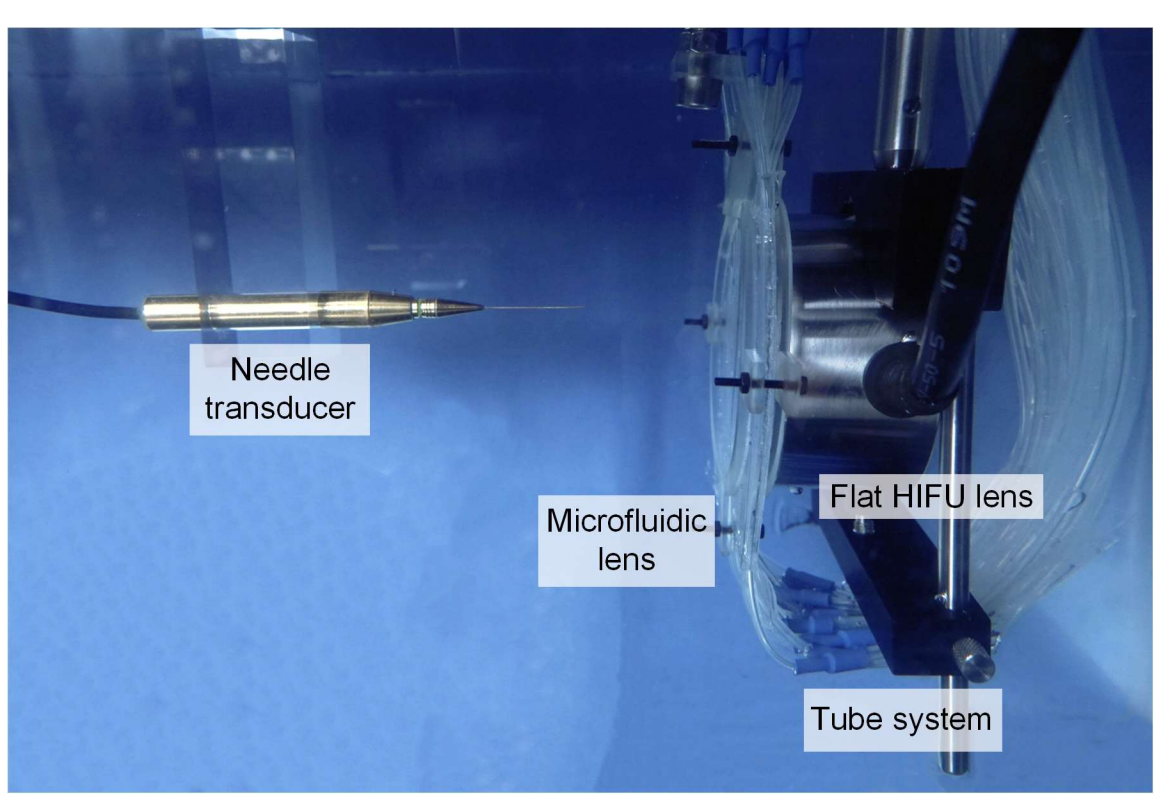

**Supplementary Fig. S18 Experiment setup for ultrasound field characterization.**

During the experiment measurement for the filed generated by the row-column microfluidic lens, a large-aperture planar high-intensity focused ultrasound (HIFU) transducer (T60-10, Siansonic Inc.) was employed to provide 1 MHz plane-wave as the input. The source signal was generated by a function generator (AFG3022C, Tektronix Inc.) and subsequently amplified by a power amplifier (AD1006, T&C Power Conversion). The spatial acoustic field distribution was characterized using a needle hydrophone (NH0075, Precision Acoustics) mounted on a high-precision 3D translation stage (**Supplementary Fig. S18**). The transmitted signal from the transducer is an ultrasound pulse (see an example in **Supplementary Fig. S5**) and the received signal is recorded using an oscilloscope (Keysight DSOX3024t).

By moving the needle hydrophone across different 2D planes, we captured time-resolved ultrasound signals at discrete spatial coordinates. During the post-processing, the raw signals were first processed with a frequency-domain bandpass filter (0.7–1.3 MHz). To isolate the pulsed signal, a temporal window was applied to each position.

The final spatial intensity maps were generated by calculating the measured amplitude of the signal inside the time window:

$$p_s(x,y) = \frac{1}{T}\int_{t_{w0}}^{t_{w0}+T} |p_{scan}(x,y,t)|dt,$$

where $p_{scan}(x,y,t)$ is the scanned raw acoustic pressure, $T$ is time window width, and $t_{w0}$ is the start of the time window, which is related to the scanning point position $r$ by $t_{w0} = r/c_{water}$.

After getting the $p_s$, a normalization is implemented to get the $p_{norm}$: $p_{norm}(x,y) = \frac{p_s(x,y)}{max(p_s(x,y))}$.

The normalized acoustic intensities were calculated by: $I(x,y) = \frac{p_{norm}(x,y)^2}{\rho_w c_w}$.

### 7.2 Ultrasound field measurement for the cylindrical lens

The performance of the cylindrical lens was measured using a rotational scanning system. A cylindrical transducer (SMC22D20H10412, STEMiNC Inc.) was placed at the center as the acoustic source, while another transducer (V303-SU, Olympus Corp.) served as the receiver. The cylindrical transducer is connected to the function generator and the power amplifier by dual conductor wires, which is also covered by a clear epoxy resin layer to isolate it with water. Unlike the linear scanning system where the sensor is moving to capture the signal, the cylindrical transducer, integrated with the microfluidic lens, was mounted on a rotational mechanical stage, while the sensing transducer remained stationary. During the rotation of the cylindrical transducer, distance between the source and sensor was fixed at 10 mm.

### 7.3 Dynamic acoustic particle manipulation

For dynamic acoustic manipulation of particles, the ultrasound transducer and the microfluidic lens were immersed in water and oriented upward. A small silicone rubber particle (cylindrical, 1.5 mm in diameter and 1 mm in height) was placed on the water surface. When the transducer was turned on, the particle was trapped at the acoustic focal spot. By dynamically reconfiguring the liquid filling patterns in the microfluidic lens to shift the focal spot position, the particle could be controllably levitated at different position across the water surface.

## 7.4 Thermal images and videos

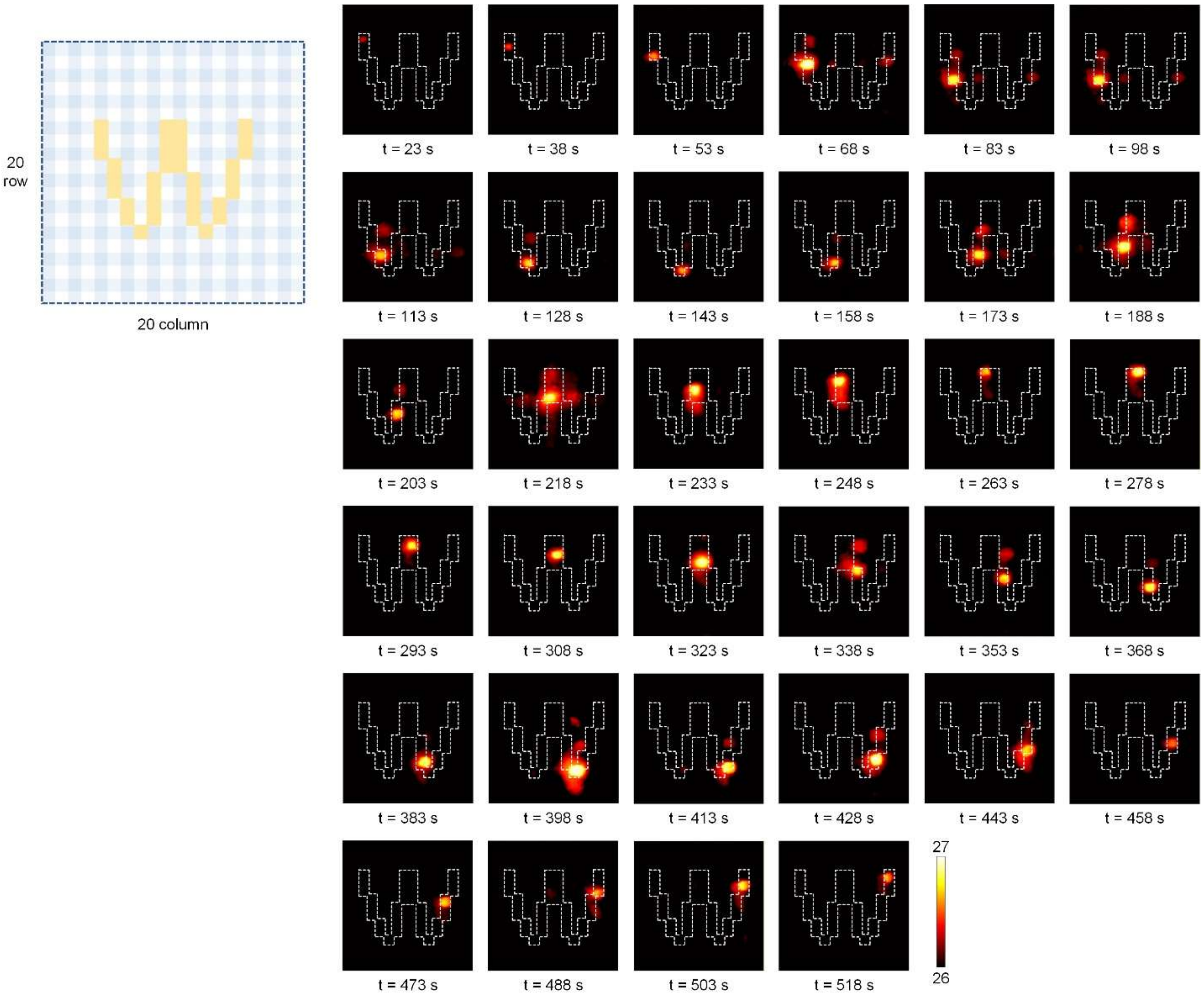

**Supplementary Fig. S19 Target heating positions and corresponding temperature distributions over time measured using a thermochromic film.**

Thermal images and videos were captured using a thermal camera (HP96, FSFTOOLS). As illustrated in **Fig. 4** in the main text, the transducer and the microfluidic lens were immersed in water and oriented upward, with a 0.5-mm-thick thermochromic film positioned at the water surface. The distance between the microfluidic lens and the water surface was set as the target focal distance of the lens. The thermochromic film was fabricated by dispersing thermochromic pigment (Lake Blue, Renfio) within a silicone rubber matrix (PIXISS) at a 1:1 mass ratio. The thermal camera was positioned directly above the film to record the temperature variation induced by the transducer and the microfluidic lens.

## Supplementary References

## Description of supplementary videos

1. **Filling pattern switch inside a single-layer lens**
2. **Filling pattern switch inside a double-layer row-column lens**
3. **Fluid switching inside the Y-shaped connector**
4. **Dynamic particle manipulation**
5. **Progressive formation of a W-shaped pattern via programmable ultrasonic heating (infrared thermal video)**
6. **Dynamic ultrasonic heating pattern transition between "V" and "W" (infrared thermal video)**